***Target enrichment of ultraconserved elements from arthropods provides a genomic perspective on relationships among Hymenoptera***

Brant C. Faircloth[1,2,*]; Michael G. Branstetter[3]; Noor D. White[4,5]; Seán G. Brady[3]

[1] Department of Ecology and Evolutionary Biology, University of California, Los Angeles, CA 90095, USA

[2] Department of Biological Sciences, Louisiana State University, Baton Rouge, LA 70803, USA

[3] Department of Entomology, National Museum of Natural History, Smithsonian Institution, Washington, D.C., 20560, USA

[4] Department of Biology, University of Maryland, College Park, MD 20742, USA

[5] Department of Vertebrate Zoology, National Museum of Natural History, Smithsonian Institution, Washington, D.C., 20560, USA.

* To whom correspondence should be addressed: brant@faircloth-lab.org; Department of Biological Sciences, Louisiana State University, Baton Rouge, LA 70803, USA

**RH**: Enrichment of UCEs from Hymenoptera

**Abstract**

Gaining a genomic perspective on phylogeny requires the collection of data from many putatively independent loci collected across the genome. Among insects, an increasingly common approach to collecting this class of data involves transcriptome sequencing, because few insects have high-quality genome sequences available; assembling new genomes remains a limiting factor; the transcribed portion of the genome is a reasonable, reduced subset of the genome to target; and the data collected from transcribed portions of the genome are similar in composition to the types of data with which biologists have traditionally worked (e.g., exons). However, molecular techniques requiring RNA as a template, including transcriptome sequencing, are limited to using very high quality source materials, which are often unavailable from a large proportion of biologically important insect samples. Recent research suggests that DNA-based target enrichment of conserved genomic elements offers another path to collecting phylogenomic data across insect taxa, provided that conserved elements are present in and can be collected from insect genomes. Here, we identify a large set (n=1,510) of ultraconserved elements (UCEs) shared among the insect order Hymenoptera. We used *in silico* analyses to show that these loci accurately reconstruct relationships among genome-enabled Hymenoptera, and we designed a set of RNA baits (n=2,749) for enriching these loci that researchers can use with DNA templates extracted from a variety of sources. We used our UCE bait set to enrich an average of 721 UCE loci from 30 hymenopteran taxa, and we used these UCE loci to reconstruct phylogenetic relationships spanning very old (≥ 220 MYA) to very young (≤ 1 MYA) divergences among hymenopteran lineages. In contrast to a recent study addressing hymenopteran phylogeny using transcriptome data, we found ants to be sister to all remaining aculeate lineages with complete support, although this result could be explained by factors such as taxon sampling. We discuss this approach and our results in the context of elucidating the evolutionary history of one of the most diverse and speciose animal orders.

## Introduction

The insect order Hymenoptera (ants, bees, wasps, and sawflies) is one of the most diverse animal orders (Grimaldi & Engel 2005), including at least 125,000 species (Gaston 1991; LaSalle & Gauld 1993a; Gaston *et al.* 1996; Sharkey 2007) and comprising approximately eight percent of all described animals (Davis *et al.* 2010). In addition to being hyperdiverse, hymenopterans are also incredibly abundant: ants alone occur in almost all terrestrial habitats and may constitute 15-20% of animal biomass in tropical rainforests. Other aculeate groups such as vespid wasps (hornets and yellow jackets) and stingless honey bees may account for an additional 15-20% (Fittkau & Klinge 1973). The ecological roles held by hymenopterans are diverse and include predator, prey, scavenger, parasite, ectosymbiont, and mutualist. Species within the order also play critical roles in worldwide pollination of agricultural crops and natural vegetation (Roubik 1995; Kevan 1999; Michener 2007), tropical forest dynamics (Roubik 1989; Levey & Byrne 1993; Dalling *et al.* 1998), ecosystem services (Kremen *et al.* 2002; Del Toro *et al.* 2012), and biological control of pest insects (Quicke 1997). Outside of their biological importance, hymenopteran taxa are models for understanding the evolution and genetic basis of social behavior (Bourke & Franks 1995; Smith et al. 2008; Bradley *et al.* 2009; Johnson & Linksvayer 2010, Howard & Thorne 2010, Wang et al. 2013), mechanisms of sex determination (Hunt & Page 1994; Beye *et al.* 1994; Page *et al.* 2002), evolution of adaptive specialization (Mueller *et* *al.* 2005; Schultz & Brady 2008; Mueller & Rabeling 2008), and origins and maintenance of biodiversity (reviewed in LaSalle & Gauld 1993b).

Given their ubiquity, diversity, biological significance, and importance to ecological and agricultural systems, resolving evolutionary relationships among Hymenoptera is critical – from the deepest splits (~220-300 MYA) within the Hymenoptera tree (Grimaldi & Engel 2005; Ronquist *et al.* 2012) to moderately deep divergences (~120-60 MYA) comprising key events in the evolution of both the ecologically dominant ants [the "Dynastic-Succession" hypothesis of (Wilson & Hölldobler 2005)] and pollinating bees (Danforth *et al.* 2013) to the very shallow divergences among lineages that may be undergoing ecological (Savolainen & Vepsäläinen 2003) or symbiont-driven speciation (Mehdiabadi *et al.* 2012). Prior molecular phylogenetic studies have made significant advances towards resolving the relationships between higher-level taxonomic groups (Sharkey 2007; Pilgrim *et al.* 2008; Heraty *et al.* 2011; Debevec *et al.* 2012; Klopfstein *et al.* 2013) and elucidating taxonomic relationships among species at shallower levels (reviewed in Moreau 2009; Danforth *et al.* 2013). However, these studies have been limited to analyzing a relatively small number of nuclear or mitochondrial loci (e.g. Brady *et al.* 2006; Danforth *et al.*, 2006; Sharanowski *et al.* 2010) that sample a small fraction of the genome.

Phylogenomic projects, such as the 1KITE initiative (http://www.1kite.org), seek to remedy this shortfall by identifying orthologous loci from widespread transcriptome sequencing. Although this approach has proven effective within Hymenoptera (Johnson *et al.* 2013), RNA-based techniques, on their own, limit the source materials useable for phylogenetic inference to fresh or properly preserved tissue (e.g. tissues stored in liquid nitrogen or RNAlater). This restriction leaves the majority of insect specimens unusable, especially those materials found in museum collections, posing a significant challenge for studies requiring rarely collected species. Thus, a significant challenge that remains for hymenopteran phylogenetics is to identify a large suite of universal markers that can be applied to samples stored with minimal preservation while maintaining the capability to elucidate relationships among lineages across a diversity of timescales.

Recent research among vertebrates has shown that target enrichment of highly conserved genomic sequences or "ultraconserved elements" (UCEs; Faircloth *et al.* 2012) provides one mechanism for meeting this challenge. UCEs are an ideal marker for phylogenetic studies as a result of their ubiquity among taxonomic groups (Siepel *et al.* 2005), low paralogy (Derti *et al.* 2006), and low saturation (McCormack *et al.* 2012). While we still do not understand the evolutionary forces driving the conservation of UCEs (Harmston *et al.* 2013) or their biological function (Bejerano *et al.* 2004; Sandelin et al. 2004; Ahituv *et al.* 2007), target enrichment of UCE loci has been used to investigate several outstanding phylogenetic questions at "deep" timescales across diverse groups of vertebrate taxa (Crawford *et al.* 2012; McCormack *et al.* 2013; Faircloth *et al.* 2013). The technique is also useful for understanding shallower, population-level events including recent divergences (Smith *et al.* 2014). When combined with massively parallel sequencing, the scalability of the UCE approach allows researchers to: parallelize the collection of data from hundreds or thousands of orthologous loci across hundreds of taxa using stable DNA inputs in a single sequencing run; reduce the data analysis burden relative to what is required for the sequencing, assembly, and alignment of multiple genomes, and conduct studies at a reasonable cost per individual.

Although enriching conserved loci resolves relationships among vertebrates, the utility of this approach among other animals is unknown. Here, we report the identification of a suite of ~1,500 UCE loci useful for inferring phylogenetic relationships across the entire Hymenoptera order. We used an *in silico* analysis to show that UCE loci recover the expected relationships among extant, genome-enabled, hymenopteran taxa with high support. We then synthesized a bait (i.e. probe) set for targeted enrichment of UCE loci, and we used the bait set to enrich an average of 721 loci among 30 sequence-tagged genomic libraries prepared from a diverse group of hymenopteran DNA sources, some of which were minimally preserved in ethanol for more than 12 years (Supplemental Table 1). Using contigs assembled from massively parallel sequencing reads of these enriched libraries, we inferred the evolutionary relationships among hymenopteran taxa spanning very deep ($\geq$ 220 MYA; estimated age of crown-group Hymenoptera; Grimaldi & Engel 2005; Ronquist *et al.* 2012) to very shallow ($\leq$ 1 MYA; estimated age of included *Nasonia* species; Werren *et al*. 2010) divergences, and we discuss our findings relative to both phylogenomic and traditional efforts to resolve the hymenopteran phylogeny.

## Materials and Methods

### *Identification of UCEs*

To identify a large set of UCE loci shared among Hymenoptera, we used LASTZ (Harris 2007) and programs from the UCE-PROBE-DESIGN package (UPDP) (https://github.com/faircloth-

lab/uce-probe-design). We aligned repeat-masked (Smit & Hubley 1996) genome assemblies of *Apis mellifera* (*apiMel4*; Honeybee Genome Sequencing Consortium 2006) and *Nasonia vitripennis* (*nasVit2*; Werren *et al.* 2010) using LASTZ. Following sequence alignment, we used rename_maf.py from UPDP to annotate the resulting multiple alignment format (MAF) lines with each taxon name. Following annotation, we used summary.py to search the resulting MAF file for aligned regions longer than 40 base pairs that were 100% conserved. We identified 2,906 conserved regions meeting these criteria, and we filtered these regions for duplicate hits using an additional LASTZ alignment of conserved regions back to themselves (all-to-all) followed by removal of matches that were more than 80% identical over 50% of their length. After removing these duplicate-like regions, we output a file of 1,555 non-duplicated UCE loci, and we checked for detection of these loci in two additional hymenopteran genome assemblies (*Atta cephalotes*, *Solenopsis invicta*; Supplemental Table 2) by aligning the conserved regions to the assemblies using LASTZ, requiring 80% sequence identity over 80% of the non-duplicate UCE locus length. Approximately 1,000-1,300 of these UCE loci were conserved across the hymenopteran genome assemblies we checked, suggesting that the suite of non-duplicated, highly conserved loci we identified were also conserved in other hymenopteran lineages.

Based on this positive result, we sliced all of the non-duplicate UCE regions from the *nasVit2* genome sequence using match coordinates (as Browser Extensible Data or BED files) output by LASTZ, and we buffered shorter UCE regions to 180 bp by including an equal amount of 5' and 3' flanking sequence from the *nasVit2* genome assembly. This buffering process allowed us to tile 120 nucleotide enrichment baits across the desired target regions at 2X tiling density (i.e., baits overlap by 60 bp; Tewhey *et al.* 2009) using py_tiler.py from the UPDP. This program also removed any resulting baits containing ambiguous base calls, having a large proportion (> 25%) of repetitive sequence, or having a high GC content (> 70%). We screened the resulting bait sequences against themselves to remove duplicate baits from the set that sometimes resulted from slicing longer, unique UCE loci into smaller, 120 nucleotide chunks. We refer to this final set of baits as the "UCE bait set" below.

*In Silico Test of UCEs*

We performed an *in silico* test of the ability of the UCE baits and their target UCE loci to resolve the phylogeny of Hymenoptera by aligning the UCE bait set to 14 hymenopteran genome assemblies downloaded from NCBI (Supplemental Table 2) using a parallel wrapper around LASTZ (run_multiple_lastzs_sqlite.py) from the PHYLUCE (https://github.com/faircloth-lab/phyluce) package. Although genome assemblies exist for additional hymenopteran taxa, we were not granted permission to include these data in our analyses. Following alignment, we sliced the UCE loci from each genome and retained ± 1000 bp of flanking sequence to the 5' and 3' end of each UCE using slice_sequence_from_genomes2.py. This program makes a first pass at removing duplicate hits during the slicing process. After slicing, and to identify assembled contigs representing UCE loci from each species using the standard PHYLUCE pipeline, we aligned species-specific UCE slices to a FASTA file of all enrichment baits using match_contigs_to_loci.py from the PHYLUCE package. This program implements the matching process using LASTZ and ensures that matches are 80% identical over 80% of their length. This program also screens and removes apparent duplicate contigs or contigs that are hit by baits targeting more than one UCE locus. After screening and removing non-target and duplicated or misassembled contigs, the program creates a relational database containing two tables – one that holds the status of each UCE locus in each taxon (detected/non-detected) and another that maps the contig names generated by the assembler to the names of the corresponding UCE locus across all taxa.

We created a file containing the names of 14 genome enabled taxa (Supplemental Table 2), and we input this list to an additional program (get_match_counts.py) that queries the relational database described above to generate a list of UCE loci shared among taxa. We input the list of loci generated by this program to another program (get_fastas_from_match_counts.py) to create a monolithic FASTA file containing all UCE sequence data for all taxa. We separated the FASTA file of sliced sequences by locus and aligned all loci using a parallel wrapper (seqcap_align_2.py) around MAFFT (version 7.130, Katoh *et al.* 2005). Following MAFFT alignment, we removed the locus names from all alignments, edge- and internally trimmed resulting alignments using the trimal "-automated1" algorithm (Capella-Gutierrez *et al.* 2009), converted trimmed alignments back to nexus format (convert_one_align_format_to_another.py), and selected the subset of alignments (get_only_loci_with_min_taxa.py) that were 70% complete (those that contained alignment data from at least 10 out of 14 taxa). We generated alignment statistics and computed the number of informative sites across all alignments using get_align_summary_data.py and get_informative_sites.py. We concatenated the resulting alignments into a PHYLIP-formatted supermatrix (format_nexus_files_for_raxml.py), we conducted 20 maximum-likelihood (ML) searches for the phylogenetic tree that best fit the data using the unpartitioned supermatrix, RAxML (version 8.0.19, Stamatakis 2006), and the GTRGAMMA model. Following the best tree search, we used RAxML to generate 100 non-parametric bootstrap replicates, we tested bootstrap replicates for convergence, and we reconciled the best fitting ML tree with the bootstrap replicates, all using features of RAxML.

*Library Preparation, Target Enrichment, and Sequencing of UCEs*

Following the *in silico* test of the UCE bait set, we had probes commercially synthesized as an RNA target capture array ("MYBaits", MYcroarray, Inc.). We then extracted DNA from 30 hymenopteran species (Supplemental Table 1) using either DNeasy extraction kits (Qiagen, Inc.) or phenol-chloroform (Maniatis *et al.* 1982) extraction procedures. We selected taxa for extraction and library preparation that span a range of divergence dates (≥ 220 MYA to < 5 MYA) and that represent major divisions within the order (sawflies, parasitoid wasps, and stinging wasps). Following extraction we quantified DNA for each sample using a Qubit fluorometer (Life Technologies, Inc.), we randomly sheared 69-509 ng (400 ng mean) DNA to a target size of approximately 650 bp (range 400-800 bp) by sonication (Qsonica Inc. Q800 or Diagenode BioRuptor), and we input the sheared DNA into a modified genomic DNA library preparation protocol (Kapa Biosystems) that incorporated "with-bead" cleanup steps (Fisher *et al.* 2011) using a generic SPRI substitute ((Rohland & Reich 2012); hereafter SPRI). This protocol is similar to the Kapa Biosystems protocol that uses commercial SPRI chemistry for cleanup and includes end-repair, adenylation, and T/A ligation steps, except that the Fisher modification does not remove and replace SPRI beads between each step. Rather, the with-bead protocol removes and replaces a 25 mM NaCl + PEG solution, leaving the beads in-solution throughout the library preparation steps until their removal just prior to PCR amplification of the library. During adapter ligation, we also substituted custom-designed sequence-tagged adapters to the ligation reaction (Faircloth & Glenn 2012). Following adapter ligation, we PCR amplified 50% of the resulting library volume (~15 µL; 50-400 ng) using a reaction mix of 25 µL HiFi HotStart polymerase (Kapa Biosystems), 5 µL of Illumina TruSeq primer mix (5 µM each), and 5 µL double-distilled water (ddH$_2$0) using the following thermal protocol: 98 C for 45s; 10-12 cycles of 98 C for 15s, 60 C for 30s, 72 C for 60s; and a final extension of 72 C for 5m. We purified resulting reactions using 1X SPRI, and we re-hydrated libraries in 33 µL ddH$_2$O. We quantified 2 µL of each library using a Qubit fluorometer. We combined groups of six libraries at equimolar ratios into enrichment pools having a final concentration of 147 ng/µL.

We prepared Cot-1 DNA from nest collections of several ant species (*Aphaenogaster fulva, Aphaenogaster rudis,* and *Formica subsericea*) following the protocol of Timoshevskiy *et al.* (2012). We followed library enrichment procedures for the MYcroarray MYBaits kit (Blumenstiel *et al.* 2010), with three modifications: (1) we added 100 ng MYBaits to each reaction (a 1:5 dilution of the standard MYBaits concentration), (2) we added 500 ng custom blocking oligos designed against our custom sequence tags and using 10 inosines to block the 10 nucleotide index sequence, and (3) for a subset of the pools (3 pools, 18 samples), we tested the efficiency of our hymenopteran Cot-1 DNA by performing duplicate enrichments adding 500 ng of hymenoptera Cot-1 versus 500 ng commercially available chicken Cot-1 DNA (Applied Genetics Laboratories, Inc.). We excluded the remaining two pools from the test and used hymenoptera Cot-1 with each. We ran the hybridization reaction for 24 hours at 65 C. Following hybridization we bound all pools to streptavidin beads (MyOne C1, Life Technologies) and washed bound libraries according to a standard target enrichment protocol (Blumenstiel *et al.* 2010).

We used two different approaches for PCR recovery of the enriched libraries. For 12 of the samples (Supplemental Table 1), we followed the standard (Blumenstiel *et al.* 2010) post-enrichment approach where we dissociated enriched DNA from RNA baits bound to streptavidin-coated beads with 0.1N NaOH, followed by a five minute neutralization of NaOH using an equal volume of 1 M Tris-HCl, a 1X SPRI cleanup, and elution of the SPRI-purified sample in 30 µL of $ddH_2O$. For the remaining 18 samples, we removed the final aliquot of wash buffer following enrichment and allowed samples to dry for five minutes while sitting in a magnet stand. We removed residual buffer with sterile toothpicks. Then, we added 30 µL $ddH_20$ to each sample and proceeded directly to PCR recovery while the enriched libraries were still bound to streptavidin beads (Fisher *et al.* 2011). The streptavidin beads do not inhibit PCR and with-bead PCR recovery of enriched libraries is a faster and easier procedure. We combined either 15 µL of unbound, SPRI-purified, enriched library or 15 µL of streptavidin bead-bound, enriched library in water with 25 µL HiFi HotStart Taq (Kapa Biosystems), 5 µL of Illumina TruSeq primer mix (5 µM each), and 5 µL of $ddH_2O$. We ran PCR recovery of each library using the following thermal profile: 98 C for 45s; 16-18 cycles of 98 C for 15s, 60 C for 30s, 72 C for 60s; and a final extension of 72 C for 5m. We purified resulting reactions using 1.8X SPRI, and we re-hydrated enriched pools in 33 µL $ddH_2O$. We quantified 2 µL of each enriched pool using a Qubit fluorometer.

Following quantification of the enriched pools, we verified enrichment and compared the utility of chicken Cot-1 to hymenopteran Cot-1 by designing primers (Untergasser *et al.* 2012) to amplify seven UCE loci (Supplemental Table 3) targeted by the baits we designed. We set up a relative qPCR by amplifying two replicates of 1 ng of enriched DNA from each library at all seven loci and comparing those results to two replicates of 1 ng unenriched DNA for each library at all seven loci. We performed qPCR using a SYBR Green qPCR kit (Hoffman-LaRoche, Ltd.) on a Roche LightCycler 480. Following data collection, we computed the average of the replicate crossing point ($C_p$) values for each library at each amplicon for each Cot-1 treatment, and we computed fold-enrichment values, assuming an efficiency of 1.78 and using the formula 1.78^abs(enriched $C_p$ - unenriched $C_p$).

Following qPCR verification and selection of the library pools that showed the greatest fold-enrichment for a given Cot-1 treatment (chicken or hymenopteran), we diluted each pool to 2.5 ng/μL for qPCR library quantification. Using the diluted DNA, we qPCR quantified libraries using a library quantification kit (Kapa Biosystems) and assuming an average library fragment length of 500 bp. Based on the size-adjusted concentrations estimated by qPCR, we created two different equimolar pools of libraries at 10 nM concentration (Supplemental Table 1), and we sequenced 9-10 pmol of each pool-of-pooled libraries using two runs of paired-end, 250 bp sequencing on an Illumina MiSeq (v2; UCLA Genotyping Core Facility).

*Analysis of Captured UCE Data*

We trimmed and demultiplexed FASTQ data output by BaseSpace for adapter contamination and low-quality bases using a parallel wrapper (https://github.com/faircloth-lab/illumiprocessor) around Trimmomatic (Bolger *et al.* 2014). Following read trimming, we computed summary statistics on the data using get_fastq_stats.py from the PHYLUCE package. To assemble the cleaned reads, we generated separate data sets using wrappers around the programs Trinity [version trinityrnaseq-r2013-02-25; assemblo_trinity.py; (Marcais & Kingsford 2011; Grabherr *et al.* 2011)] and ABySS [version 1.3.6; assemblo_abyss.py; (Simpson *et al.* 2009)]. For both assemblies we computed coverage across assembled contigs by using a program (get_trinity_coverage.py) that re-aligns the trimmed sequence reads to each set of assembled contigs using BWA-MEM (Li 2013), cleans the resulting BAM files using PICARD (version 1.99; http://picard.sourceforge.net/), adds read-group (RG) information to each library using PICARD, indexes the resulting BAM file using SAMTOOLS (Li *et al.* 2009), and calculates coverage at each base of each assembled contig using GATK [version 2.7.2, (Van der Auwera *et al.* 2002; McKenna *et al.* 2010; DePristo *et al.* 2011)].

To identify assembled contigs representing enriched UCE loci from each species, we aligned species-specific contig assemblies from both sequence assembly programs to a FASTA file of all enrichment baits using match_contigs_to_loci.py, as described above. We created a file containing the names of 30 enriched taxa from which we collected data (Supplemental Table 1), as well as the names of 14 genome-enabled, hymenopteran taxa (Supplemental Table 2), and we input this list to an additional program (get_match_counts.py) that queries the relational database created by matching baits to assembled contigs, as well as the relational database containing UCE match data for genome-enabled taxa (created as part of the *in silico* tests), to generate a list of UCE loci shared among all taxa. We input the list of loci generated by this program to an additional program (get_fastas_from_match_counts.py) to create a monolithic FASTA file containing all UCE sequence data for all taxa. We aligned all data in the monolithic FASTA file using MAFFT (Katoh *et al.* 2005) and seqcap_align_2.py, as described above. Following

MAFFT alignment, we removed the locus names from all alignments (remove_locus_name_from_nexus_lines.py), edge- and internally trimmed resulting alignments using the “-automated1” algorithm implemented in trimal (Capella-Gutierrez *et al.* 2009), converted trimmed alignments back to nexus format (convert_one_align_format_to_another.py), and selected the subset of alignments (get_only_loci_with_min_taxa.py) that were 75% complete (those that contained alignment data from at least 33 out of 44 individuals). We generated alignment statistics and computed the number of informative sites across all alignments using get_align_summary_data.py and get_informative_sites.py.

We concatenated the resulting alignments into a PHYLIP-formatted supermatrix (format_nexus_files_for_raxml.py) and conducted 20 maximum-likelihood (ML) searches for the phylogenetic tree that best fit the data using the unpartitioned supermatrix, RAxML (Stamatakis 2006), and the GTRGAMMA model. Following the best tree search, we used RAxML to generate 100 non-parametric bootstrap replicates, we tested bootstrap replicates for convergence, and we reconciled the best fitting ML tree with the bootstrap replicates.

**Results**

*Identification of UCEs*

We identified 1,510 non-duplicate, 60 bp regions of 100% conservation across the alignments of *apiMel4* to *nasVit2*, and we designed a capture bait set containing 2,749 probes targeting these 1,510 loci.

*In Silico Test of UCEs*

During our *in silico* tests, we located an average of 863.7 (95 CI: 98.3) unique UCE loci across genome-enabled hymenopteran species (Supplemental Table 2). Following identification and filtering for uniqueness, sequence slicing, sequence alignment, and trimming of resulting alignments, we generated a 70% complete matrix containing 721 UCE loci and having a mean alignment length of 1,434 base pairs (95 CI: 35.5). These loci contained an average of 819 informative sites per locus, and concatenation of all loci in the complete matrix produced a supermatrix of 1,033,906 bp containing 591,033 informative sites. The phylogeny inferred from these results (Supplemental Figure 1) reconstructs the established relationships among genome-enabled hymenopteran lineages (Brady *et al.* 2006; Werren *et al.* 2010; Heraty *et al.* 2011; Oxley *et al.* 2014) with complete support.

*In Vitro Test of UCEs*

We extracted an average of 1,894 ng DNA (181–6,480 ng) from each hymenopteran species and input an average of 400 ng (69–509 ng) to the library preparation process. Following library prep, PCR amplification, and SPRI purification, DNA libraries contained approximately 100 ng DNA (53-151 ng). Fold enrichment values of enriched libraries estimated by qPCR suggested that commercial chicken Cot-1 performed better than the hymenopteran Cot-1 we prepared by roughly 500-fold (Supplemental Table 4): pooled libraries blocked with chicken Cot-1 showed an average fold enrichment of 744x while pooled libraries blocked with hymenopteran Cot-1 showed an average fold enrichment of 178x. Based on these results, we sequenced the three enriched pools where we could choose chicken Cot-1 as blocking DNA, as well as the remaining two pools where we could only choose hymenopteran Cot-1 as blocking DNA.

Sequencing produced an average of 1.1 million (95 CI: 249,342) reads per sample. Reads averaged 192.6 bp (95 CI: 3.7) following demultiplexing, quality-, and adapter-trimming (Supplemental Table 5). Using Trinity (Table 1), we assembled these DNA reads into an average of 74,140 contigs of 347.7 bp in length (95 CI: 6.9) and having a mean coverage of 4.1X (95 CI: 0.3). Using ABySS (Supplemental Table 6), we assembled these DNA reads into an average of 143,863 contigs of 202.7 bp in length (95 CI: 5.1) and having a mean coverage of 3.8X (95 CI: 0.2).

After searching for UCEs within the Trinity assemblies (Table 1; Supplemental Table 7), we enriched an average of 721 (95 CI: 48.2) unique UCE loci, the average locus length was 1,010 bp (95 CI: 66.1), the average coverage per enriched UCE locus was 52.3X (95 CI: 9.4), and the mean percentage of reads-on-target was 30% (95 CI: 2.7%). When searching against the ABySS assemblies (Supplemental Table 6, Supplemental Table 8), we enriched an average of 477 (95 CI: 56.4) unique UCE loci, the average locus length was 669.1 bp (95 CI: 36.9), the average coverage per enriched UCE locus was 40.7X (95 CI: 5.1), and the mean percentage of reads-on-target was 12.5% (95 CI: 3%). These and other summary statistics on assemblies (Supplemental Figure 2) suggest that Trinity assembled UCE contigs have more desirable properties for downstream phylogenetic analyses, in aggregate, than ABySS assemblies.

Following alignment of the Trinity-assembled data, alignment trimming, and filtering of loci having fewer than 33 taxa (75 % complete), we retained 600 alignments having an average length of 691.4 bp (95 CI: 44.4 bp). The average number of taxa present in these 600 alignments was 39.2 (95 CI: 0.2). The concatenated, Trinity supermatrix contained 414,849 bp, 413,782 total nucleotide characters, and 282,973 informative sites. Following alignment of the ABySS-assembled data, alignment trimming, and filtering of loci having fewer than 33 taxa (75 % complete), we retained 196 alignments of 522.5 bp (95 CI: 82.1 bp) in length. The average number of taxa present in these 196 alignments was 36.71 (95 CI: 0.37). The concatenated, ABySS supermatrix contained 102,418 bp, 102,148 total nucleotide characters, and 60,714 informative sites.

We inferred a phylogeny from both Trinity assemblies (Figure 1) and ABySS assemblies (Supplemental Figure 3). Because the Trinity assemblies produced a larger number of longer, higher coverage UCE loci that yielded a larger, 70% complete, concatenated supermatrix, we focus on the relationships we inferred from the Trinity data. However, the ABySS topology (Supplemental Figure 3), while having slightly lower support at several nodes, was identical to the topology we inferred from the Trinity assemblies.

Generally, the relationships among Hymenoptera we inferred from the Trinity supermatrix (Figure 1) accurately reconstructed: (1) the relationships among-genome enabled hymenopterans inferred during our *in-silico* analysis, (2) the established relationships between taxa from which we collected data, *de novo*, and (3) the established relationships between genome-enabled taxa and species from which we collected data (Danforth *et al.* 2006; Brady *et al.* 2006; Werren *et al.* 2010; Heraty *et al.* 2011; Branstetter 2012; Oxley *et al.* 2014; Ward *et al.* In press).

In Figure 1, the sawflies, represented here by only the superfamily Tenthredinoidea, formed a clade sister to the Apocrita. Within the Apocrita, parasitic wasps formed a paraphyletic grade leading to a monophyletic Aculeata (stinging wasps, ants, and bees) with *Orthogonalys* (Trigonalidae)+*Evaniella* (Evaniidae) recovered as sister to the aculeates. Within Aculeata, we recovered five main groups with maximum support (note that we did not include chrysidoid wasps): ants (Formicidae), spheciform bees+wasps (Apoidea), vespid wasps (Vespidae), scoliid wasps (Scoliidae), and tiphioid-pompiloid wasps (Chyphotidae+Pompilidae+Sapygidae). Among these groups, we inferred the ants to be sister to a clade containing all remaining aculeate lineages with maximum support. Within the clade containing the remaining aculeates, we recovered the Scoliidae as sister to the Apoidea (87% support), and we recovered the Vespidae as sister to the tiphioid-pompiloid wasps (58% support). Within the ants, we recovered all expected relationships among the five included subfamilies (Ponerinae, Dorylinae, Dolichoderinae, Formicinae, and Myrmicinae; Brady *et al.* 2006; Moreau & Bell 2013), and several closely related ant genera and species belonging to the tribe Stenammini (*Aphaenogaster*, *Messor*, and *Stenamma*; Branstetter 2012; Ward *et al.* In press) with high (≥ 87%) support.

To test the effects of removing distantly related sawfly lineages on the topology and support inferred across the UCE data, we constructed a new UCE data set lacking sawfly lineages because the sawfly data were the most incomplete, with respect to counts of recovered loci across all taxa (see Supplemental Figure 4 and below), and the inclusion of sawflies had the largest effect on the size of our incomplete matrix. This new dataset (75% complete) included 638 UCE loci, contained an average of 37.2 taxa (95 CI: 0.2), and had an average alignment length of 737.1 bp (95 CI: 46.4) The supermatrix contained 470,258 bp, 469,081 total nucleotide characters, and 310,253 (+27,280) informative sites.  Following inference from this updated dataset with RAxML using approaches identical to those described above, the resulting phylogeny (Supplemental Figure 6) had the same topology as the tree including sawflies with the exception of inferred relationships between two non-aculeate taxa, *Evaniella* and *Orthognalys*.

*Analysis of Capture Success*

Based on the differences in capture success we observed across the resulting phylogeny (Supplemental Figure 4), we analyzed several summary metrics (Supplemental Table 7, Supplemental Table 8), *post-hoc,* using general linear models (R, version 2.5.12; (Team 2011)) to investigate those parameters affecting the number (Poisson link function) and length (Gaussian link function) of UCE loci we recovered. With these values, we also included an explicit measure of pairwise genetic distance between all taxa from which we enriched sequence data and the *nasVit2* genomic assembly, from which we designed capture baits. We estimated distance values from the concatenated Trinity supermatrix using the "distance" method of PyCogent (version 1.5.3; (Knight *et al.* 2007)) and assuming a GTR site rate substitution model. We used Akaike's Information Criteria (AIC) to rank and compare linear models, and we model-averaged estimates across parameters where there was a valid set ($w_i > 0.10$; (Royall 1997)) of candidate models. These *post-hoc* analyses suggest that UCE capture success may be driven by several factors, in addition to phylogenetic distance between the probe design source and the taxa being enriched. Specifically, Akaike weights suggest that a "global" model containing four parameters (distance + reads + mean read length + assembly method) best approximates the data (Supplemental Table 9), that there are large differences among parameter effect sizes (Supplemental Figure 5), and that phylogenetic distance has the largest effect of parameters we investigated on the number of UCE contigs enriched. The size of this effect is tempered somewhat when considering only the Trinity assemblies, where read length appears to play a role (Supplemental Table 10, Supplemental Figure 5). Similarly, length of enriched UCE contigs may best be explained (Supplemental Table 11) by a global model containing 3 parameters (distance + reads + assembly method), assembly method likely plays a larger role in resulting length of UCE loci, and phylogenetic distance retains a large effect on resulting contig length (Supplemental Figure 5). When considering only the Trinity assemblies, the effects of distance and the number of reads are both important factors affecting resulting contig length (Supplemental Figure 12). In all of these results, it is important to keep in mind that phylogenetic distance falls on the interval [0,1], so the effect size of this parameter is tempered by typically small changes in its value.

**Discussion**

We have developed a powerful new genomic tool for estimating phylogenetic relationships among members of the hyperdiverse insect order Hymenoptera. By extending and improving prior work (Faircloth *et al.* 2012), we identified over 1,500 highly conserved genomic regions between distantly related Hymenoptera taxa, collected these loci from 14 genome-enabled and 30 non-genome-enabled taxa using *in silico* and *in vitro* techniques, and used the resulting genome-scale sequence data to accurately infer both deep (~220-300 MYA) and relatively shallow (≤ 1 MYA) relationships. Although other phylogenomic approaches have been employed among arthropods (Johnson *et al.* 2013), this is the first time that sequence capture of conserved regions has been used to collect genome-scale DNA data from this group.

Compared to recent phylogenetic studies investigating higher-level relationships within Hymenoptera (Sharkey 2007; Heraty *et al.* 2011; Klopfstein *et al.* 2013), the UCE data recovered all well-established relationships with complete support. In addition, the UCE data suggest a novel relationship within the Aculeata, in which the ants are sister to all remaining aculeate lineages included here. The aculeates contain all major lineages of social insects (except termites) including ants, vespid wasps, and several lineages of social bees. Aculeata also includes the most important group of pollinators (bees). Hence, understanding relationships among the aculeates is critical to provide the comparative framework needed to study the origins and evolution of sociality and pollination biology in this group (Danforth 2013). Until recently, phylogenetic studies of aculeates have been based on a relatively small number of characters and have produced conflicting results (Brothers 1999; Pilgrim *et al.* 2008; Peters *et al.* 2011; Debevec *et al.* 2012). A recent transcriptome-based study (Johnson *et al.* 2013) sequenced key lineages within Aculeata and produced a fully resolved phylogeny of aculeate lineages, recovering a novel relationship in which ants are sister to the Apoidea (spheciform bees+wasps). Our UCE data set did not recover this relationship. Instead, we found ants to be sister to all remaining aculeate lineages with complete support, but there were several nodes within each clade receiving moderate (≥ 58%) support. Our study also differed from Johnson *et al.* (2013) in the placement of vespid wasps as sister to the tiphioid-pompiloid wasps (Chyphotidae+Pompilidae+Sapygidae) and the scoliid wasps as sister to the spheciform

wasps+bees (Apoidea). Previous work by Debevec *et al.* (2012) also recovered this placement of scoliid wasps as sister to the spheciform wasps+bees.

Given the importance of resolving relationships among aculeate lineages, we tested the effects of removing sawfly lineages on the topology and support inferred across the UCE tree presented in Figure 1. Following inference from this updated dataset with RAxML, the resulting phylogeny (Supplemental Figure 6) had the same topology as the tree including sawflies, except that in Figure 1, two non-aculeate taxa, *Evaniella* and *Orthognalys* form a clade with maximum support, while in Supplemental Figure 6, these taxa form a grade, also with maximum support. Support values for internal nodes were marginally higher in the tree excluding sawflies. The stability of the recovered relationships within Aculeata between these two trees and across different assembly methods suggests that neither the count of loci, nor the total amount of data, nor the assembly approach are driving the differences we observed between our results and those of Johnson et al. (2013).

Rather, taxon sampling (e.g. our study does not include any chrysidoid wasps) or other differences among each data set including size, analytical approach, nucleotide composition, locus type, the number of independent loci sampled, and matrix completeness could explain the differences in topology we observed. For example, Johnson et al. (2013) collected and analyzed both larger and smaller amounts of data (175,404 – 3,001,657 sites) of a different type (amino acid residues) from fewer taxa (n=19) that included variable counts of loci (308 – 5,214 genes) spanning a range of matrix completeness (50-100%), and they inferred their phylogeny using concatenated maximum likelihood, concatenated Bayesian, and summary-statistic gene tree species tree approaches. In contrast, we collected and analyzed a less variable amount of data (102,418-469,081 sites), from a larger number of taxa (n=41-43) that included variable counts of loci (196 – 638 loci) spanning a small range of matrix completeness (70-75%). We inferred the phylogeny using a concatenated maximum likelihood approach. The types of differences between these two studies and their effects on phylogenetic reconstruction are the sorts of questions that deserve the bulk of current and future analytical effort in phylogenomics.

Focusing within ants, we captured an average of 748 UCE loci (95 CI: 5.0) using the bait set we designed and inferred nearly all relationships with complete support. The relationships we recovered among ant subfamilies agree with several recent molecular phylogenies of ants (Moreau *et al.* 2006; Brady *et al.* 2006; Moreau & Bell 2013). Furthermore, most relationships within the tribe Stenammini (*Aphaenogaster, Messor,* and *Stenamma*), including relationships within *Stenamma,* agree with recent molecular studies (Moreau *et al.* 2006; Brady *et al.* 2006; Branstetter 2012; Moreau & Bell 2013). Our study also agrees with a recent 11-gene phylogeny that documents the non-monophyly of the genus *Aphaenogaster* (Ward *et al.* In press). These observations are important because they demonstrate the potential for using UCEs to resolve shallow relationships within the Hymenoptera (divergence dates among *Stenamma* species are estimated at ~35 to < 5 MYA (Branstetter 2012)) similar to results from UCE data collected among vertebrates (Faircloth *et al.* 2013; Smith *et al.* 2014).

A major advantage of the UCE approach we describe over transcriptome-based methods is that it does not require specially preserved tissues. Here, we successfully extracted and enriched DNA from insect specimens that ranged from 12 years old to weeks old using a variety of collection methods, including several that were suboptimal for DNA preservation (ethanol preserved or dry pinned) and resulted in the extraction of little DNA (Supplemental Table 1). Furthermore, we successfully generated and enriched UCE loci from genomic libraries constructed using as little as 70 ng of DNA. This finding is significant because many arthropod taxa are small, yielding very low amounts of DNA, and our results suggest we can successfully prepare and enrich libraries from low DNA inputs. New library preparation approaches, including the Hyper Prep Kit (Kapa Biosystems) and the NEBNext Ultra Kit (New England Biolabs), should make it possible to use even less DNA in the future without resorting to expensive modifications of protocol. The ability to use small, moderately old, and sometimes low-quality specimens with the UCE approach we describe means that much of the available materials in museums and other collections can be used as a DNA source for phylogenomic studies - making it possible to sequence very rare and, often, very important taxa.

**Acknowledgements**

We thank Alex Buerkle, Jennifer Mandel, and three anonymous reviewers for their comments, which improved this manuscript. We thank Dave Smith, Ana Jesovnik, Jack Longino, Phil Ward, Brian Fisher, Peter Hawkes, Jim Carpenter, James Pitts, Gary Hevel, and Sam Droege for providing some of the tissue samples we used. Hong Zhao helped with DNA extraction and Alex Chase assisted with library preparation. We thank Uma Dandekar and Hemani Wijesuriya for performing MiSeq runs and Travis Glenn for helpful discussions. Phil Ward provided insightful comments regarding our results and relationships within the Aculeata. To make the cartoon Hymenoptera used in Figure 1, we altered open-access or public domain photographs taken by the following individuals: Bob Peterson (*Camponotus* sp.), Srikaanth Sekar (*Harpegnathos* sp.),

Gilles San Martin (*Stenamma* sp.), Christophe Quintin (*Taxonus* sp.), Franco Folini (*Apis* sp.), M.E. Clark (*Nasonia* sp.), Muhammad Mahdi Karim (Cremastinae), OpenCage (Vespidae), and DeadStar (*Acromyrmex* sp.). All modified images are available as PSD files from the Dryad link, below.

**Data and Source Code Availability**

The bait set and enrichment protocols used as part of this manuscript are available under Creative Commons license (CC-BY-3.0) from http://ultraconserved.org. The bait set is also available from Dryad (doi: 10.5061/dryad.46195). Computer programs used throughout this study are available from https://github.com/faircloth-lab/uce-probe-design, http://github.com/faircloth-lab/phyluce, and https://github.com/faircloth-lab/illumiprocessor under an open-source, BSD-style license. Sequence reads generated as part of this manuscript are available from the NCBI Sequence Read Archive (SRA PRJNA248919). Trinity contig assemblies representing UCE loci are available from Genbank (KM411995-KM433620). Additional data including the probe set design file, sequence alignments, alignment supermatrices, configuration files, estimated phylogenetic distances, qPCR results, and inferred trees are also available from Dryad (doi: 10.5061/dryad.46195).

**Support**

A Smithsonian Institution grant from the Consortium for Understanding and Sustaining a Biodiverse Planet (to SGB, BCF, and NDW) provided most of the funding for this project and support for NDW's participation. A Smithsonian Institution Buck Postdoctoral Fellowship and NSF grants DEB-1354739 and DEB-1354996 (project ADMAC) supported MGB. NSF DEB-1242260 (to BCF) and an Amazon Web Services Education Grant (to BCF) supported computational portions of this work. NSF EF-0431330 (to SGB) provided resources to obtain some specimens used in this study. This study was also supported in part by resources and technical expertise from the Georgia Advanced Computing Resource Center, a partnership between the University of Georgia's Office of the Vice President for Research and Office of the Vice President for Information Technology.

**Attributions**

SGB, BCF, and NDW contributed text to the grant application supporting this work. SGB, BCF, MGB, and NDW designed the study. BCF, NDW, MGB conducted the laboratory work. BCF analyzed the data. BCF, MGB, SGB wrote the manuscript. All authors discussed the results and commented on the manuscript.

## Figure Legends

**Figure 1**: Maximum likelihood phylogeny inferred from a 75% complete supermatrix containing data from 14 genome-enabled taxa (identified by double-asterisks) and 30 taxa from which we enriched and assembled (Trinity) ultraconserved element loci. We show bootstrap support values only where support is < 100%, and the single asterisk beside *Stenamma megamanni* denotes that this sample represents a different population of the same species.

**Table 1**: Summary values describing the number of contigs assembled by Trinity from adapter- and quality-trimmed reads ("All" contigs), their average coverage, the mean length of All contigs, the count of unique reads aligned to All contigs, the number of UCE contigs identified from the pool of All contigs, the mean length of UCE contigs, the average UCE contig sequencing coverage, and the percentage of unique reads that aligned to UCE contigs (this is a percentage of the percentage of unique reads aligning to All contigs).

| Taxon | All contigs | All contigs coverage | All contigs coverage 95 CI | All contigs mean length | All contigs mean length 95 CI | All contigs unique reads aligned | UCE contigs | UCE contigs mean length | UCE contigs coverage | UCE contigs unique reads aligned |
|---|---|---|---|---|---|---|---|---|---|---|
| *Acordulecera pellucida* | 30,033 | 3.4 | 0.2 | 344.9 | 2.6 | 70.7% | 341 | 1,025.0 | 26.3 | 18.4% |
| *Andrena (Callandrena) asteris* | 4,587 | 4.6 | 0.3 | 358.0 | 6.2 | 69.2% | 740 | 574.7 | 9.8 | 44.4% |
| *Andrena (Melandrena) sp* | 33,761 | 3.3 | 0.3 | 345.1 | 3.4 | 68.5% | 774 | 857.0 | 18.2 | 25.9% |
| *Aphaenogaster albisetosa* | 157,813 | 3.9 | 0.1 | 359.4 | 1.0 | 78.7% | 764 | 1,128.6 | 88.3 | 26.0% |
| *Aphaenogaster megommata* | 117,378 | 4.5 | 0.3 | 341.6 | 1.2 | 76.6% | 751 | 1,184.1 | 79.2 | 28.8% |
| *Aphaenogaster tennesseensis* | 76,656 | 4.8 | 0.1 | 336.8 | 1.7 | 67.1% | 751 | 1,056.3 | 62.4 | 30.5% |
| *Aphaenogaster texana* | 49,221 | 4.9 | 0.1 | 330.3 | 2.3 | 68.9% | 750 | 924.6 | 51.5 | 33.2% |
| *Aporus niger* | 16,552 | 6.3 | 2.7 | 332.5 | 3.8 | 68.2% | 740 | 714.1 | 14.5 | 17.7% |
| *Bombus pensylvanicus* | 26,877 | 3.2 | 0.1 | 321.4 | 2.7 | 72.2% | 780 | 859.7 | 21.4 | 35.0% |
| *Chalybion californicus* | 44,727 | 3.8 | 0.6 | 324.1 | 1.4 | 69.0% | 778 | 812.2 | 33.2 | 29.6% |
| *Chyphotes mellipes* | 105,477 | 4.4 | 0.7 | 383.5 | 1.6 | 79.3% | 774 | 1,184.0 | 66.2 | 26.6% |
| *Evaniella semaeoda* | 18,980 | 4.5 | 0.2 | 359.1 | 4.3 | 73.6% | 638 | 971.7 | 31.1 | 39.1% |
| *Messor piceus* | 91,858 | 4.6 | 0.8 | 332.6 | 1.4 | 71.7% | 730 | 1,111.7 | 58.9 | 26.2% |
| *Metapolybia cingulata* | 63,299 | 3.3 | 0.1 | 326.0 | 1.3 | 77.0% | 685 | 823.3 | 40.1 | 24.7% |
| *Mischocyttarus flavitarsis* | 16,624 | 5.5 | 1.6 | 330.2 | 3.6 | 82.1% | 634 | 711.2 | 30.0 | 32.4% |
| *Nasonia vitripennis* | 27,195 | 4.9 | 0.2 | 314.2 | 2.1 | 77.2% | 1166 | 771.1 | 46.9 | 57.1% |
| *Nematus tibialis* | 48,874 | 3.5 | 0.1 | 350.3 | 2.1 | 72.4% | 453 | 1,049.5 | 47.9 | 26.4% |
| *Orthogonalys pulchella* | 106,246 | 4.1 | 0.1 | 405.2 | 1.4 | 87.9% | 706 | 1,364.0 | 109.0 | 35.0% |
| *Pogonomyrmex occidentalis* | 154,514 | 3.9 | 0.1 | 362.2 | 1.0 | 83.5% | 741 | 1,142.4 | 97.5 | 26.8% |
| *Sapyga pumila* | 108,990 | 4.0 | 0.1 | 361.6 | 1.4 | 77.5% | 720 | 1,046.9 | 86.4 | 28.6% |
| *Scolia verticalis* | 55,545 | 3.9 | 0.3 | 350.7 | 1.9 | 75.5% | 760 | 1,070.4 | 56.6 | 36.0% |
| *Sericomyrmex harekulli* | 25,698 | 3.5 | 0.1 | 329.0 | 2.6 | 71.3% | 744 | 814.8 | 22.3 | 33.5% |
| *Stenamma diecki* | 108,642 | 3.9 | 0.1 | 365.8 | 1.7 | 71.8% | 751 | 1,142.0 | 53.5 | 23.7% |
| *Stenamma expolitum* | 135,131 | 3.7 | 0.1 | 363.2 | 1.3 | 76.9% | 749 | 1,212.1 | 69.3 | 25.7% |
| *Stenamma felixi* | 138,761 | 3.8 | 0.1 | 341.7 | 1.1 | 77.7% | 762 | 1,071.8 | 75.3 | 25.1% |
| *Stenamma impar* | 89,581 | 4.4 | 0.3 | 355.3 | 2.1 | 68.7% | 741 | 1,056.0 | 49.8 | 22.4% |
| *Stenamma megamanni* | 78,363 | 3.0 | 0.0 | 354.6 | 1.6 | 75.8% | 754 | 1,138.0 | 37.8 | 28.6% |
| *Stenamma megamanni2* | 147,772 | 3.8 | 0.1 | 359.5 | 1.3 | 79.7% | 756 | 1,232.9 | 87.5 | 28.7% |
| *Stenamma muralla* | 102,541 | 3.4 | 0.1 | 334.9 | 1.1 | 79.7% | 734 | 1,132.0 | 61.6 | 30.6% |
| *Taxonus pallidicornis* | 42,507 | 3.3 | 0.1 | 356.4 | 2.5 | 71.6% | 459 | 1,140.9 | 37.7 | 27.5% |

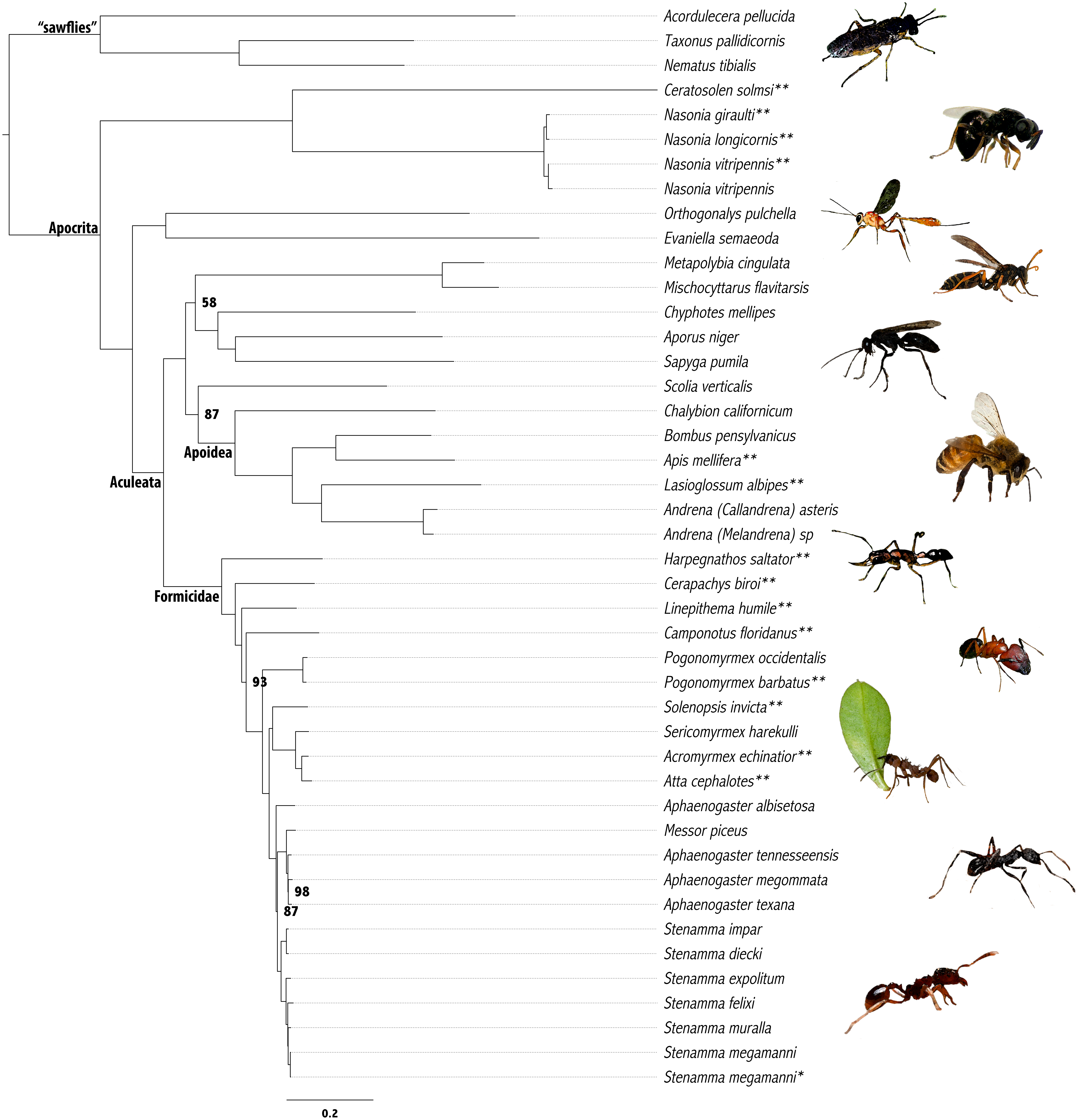

**Supplemental Table 1**: Family, species, collection identifier, collection year, collection country, collection method, voucher identifier, voucher depository, total amount of extract DNA, amount of DNA input to library preparation, post-enrichment method, and MiSeq run of all samples used for target enrichment.

| Family | Species | CollnID | CollnYr | Country | CollnMethod | VoucherID | Deposited | DNA (ng) | Lib DNA (ng) | Post enrichment method | MiSeq Run |
|---|---|---|---|---|---|---|---|---|---|---|---|
| Andrenidae | *Andrena (Callandrena) asteris* | USGS_DRO_137659 | 2009 | USA | hand net | USNMENT00921242 | USNM | 6480 | 486 | with bead | R1 |
| Andrenidae | *Andrena (Melandrena) sp* | None | 2012 | USA | Malaise trap | USNMENT00921243 | USNM | 693 | 69 | with bead | R1 |
| Apidae | *Bombus pensylvanicus* | USGS_DRO_13724 | 2009 | USA | soapy pan trap | USNMENT00921244 | USNM | 560 | 470 | with bead | R1 |
| Bradynobaenidae | *Chyphotes mellipes* | PSW14654 | 2002 | USA | unknown | CASENT0106101 | CASC | 678 | 498 | with bead | R1 |
| Evaniidae | *Evaniella semaeoda* | D. Smith 5 | 2012 | USA | Malaise trap | USNMENT00921245 | USNM | 2340 | 503 | with bead | R1 |
| Formicidae | *Aphaenogaster albisetosa* | MGB1978 | 2011 | USA | hand collection | USNMENT00921246 | USNM | 784 | 500 | NaOH | R2 |
| Formicidae | *Aphaenogaster megommata* | PSW16689 | 2012 | Mexico | hand collection | USNMENT00921247 | USNM | 319 | 266 | NaOH | R2 |
| Formicidae | *Aphaenogaster tennesseensis* | MGB2040 | 2013 | USA | hand collection | USNMENT00921248 | USNM | 779 | 500 | NaOH | R2 |
| Formicidae | *Aphaenogaster texana* | MGB1983 | 2011 | USA | litter sifting | USNMENT00921249 | USNM | 736 | 500 | NaOH | R2 |
| Formicidae | *Messor piceus* | TRP-2012a-NPHC(B) | 2012 | S. Africa | hand collection | USNMENT00921250 | USNM | 932 | 500 | NaOH | R2 |
| Formicidae | *Pogonomyrmex occidentalis* | MGB2005 | 2012 | USA | hand collection | USNMENT00921251 | USNM | 1021 | 500 | NaOH | R2 |
| Formicidae | *Sericomyrmex harekulli* | AJ111125-05 | 2011 | Guyana | bait | USNMENT00921252 | USNM | 843 | 497 | with bead | R1 |
| Formicidae | *Stenamma diecki* | MGB2046 | 2013 | USA | hand collection | USNMENT00921253 | USNM | 284 | 237 | NaOH | R2 |
| Formicidae | *Stenamma expolitum* | MGB1880 | 2011 | Nicaragua | hand collection | USNMENT00921254 | USNM | 400 | 333 | NaOH | R2 |
| Formicidae | *Stenamma felixi* | JTL7524 | 2011 | Nicaragua | hand collection | USNMENT00921255 | USNM | 533 | 444 | NaOH | R2 |
| Formicidae | *Stenamma impar* | MGB2036 | 2013 | USA | hand collection | USNMENT00921256 | USNM | 262 | 218 | NaOH | R2 |
| Formicidae | *Stenamma megamanni* | Wm-D-05-1-01 | 2011 | Nicaragua | litter sifting | USNMENT00921257 | USNM | 315 | 252 | with bead | R1 |
| Formicidae | *Stenamma megamanni* | MGB1764 | 2011 | Nicaragua | hand collection | USNMENT00921258 | USNM | 326 | 272 | NaOH | R2 |
| Formicidae | *Stenamma muralla* | JTL7007 | 2010 | Honduras | hand collection | USNMENT00921259 | USNM | 181 | 151 | NaOH | R2 |
| Pergidae | *Acordulecera pellucida* | D. Smith 3 | 2012 | USA | Malaise trap | USNMENT00921260 | USNM | 1060 | 424 | with bead | R1 |
| Pompilidae | *Aporus niger* | ANTC4004 | 2002 | USA | unknown | CASENT0106104 | CASC | 2120 | 498 | with bead | R1 |
| Pteromalidae | *Nasonia vitripennis*** | None | 2013 | USA | purchased | None | None | 3923 | 500 | with bead | R1 |
| Sapygidae | *Sapyga pumila* | ANTC4005 | 2004 | USA | unknown | CASENT0106105 | CASC | 2270 | 499 | with bead | R1 |
| Scoliidae | *Scolia verticalis* | ANTC4007 | 2004 | Australia | unknown | CASENT0106107 | CASC | 5550 | 500 | with bead | R1 |
| Sphecidae | *Chalybion californicum* | PSW15440 | 2005 | USA | unknown | CASENT0106103 | CASC | 3730 | 506 | with bead | R1 |
| Tenthredinidae | *Nematus tibialis* | D. Smith 2 | 2012 | USA | Malaise trap | USNMENT00921261 | USNM | 2950 | 502 | with bead | R1 |
| Tenthredinidae | *Taxonus pallidicornis* | D. Smith 1 | 2012 | USA | Malaise trap | USNMENT00921262 | USNM | 6120 | 490 | with bead | R1 |
| Trigonalidae | *Orthogonalys pulchella* | D. Smith 4 | 2012 | USA | Malaise trap | USNMENT00921263 | USNM | 3390 | 509 | with bead | R1 |
| Vespidae | *Metapolybia cingulata* | ANTC4006 | 2007 | Peru | unknown | CASENT0106106 | USNM | 749 | 502 | with bead | R1 |
| Vespidae | *Mischocyttarus flavitarsis* | PSW15442 | 2005 | USA | unknown | CASENT0106102 | USNM | 6480 | 486 | with bead | R1 |

**Supplemental Table 2**: Species, genome assembly, genome assembly source, reference, and number of UCE loci in assembly for all genome-enabled taxa.

| **Species** | **Name of assembly** | **Assembly source** | **Reference** | **Unique UCE loci in assembly** |
|---|---|---|---|---|
| *Acromyrmex echinatior* | Aech_3.9 | Genbank | 10.1101/gr.121392.111 | 774 |
| *Apis mellifera* | Amel_4.5 | Genbank | 10.1186/1471-2164-15-86 | 803 |
| *Atta cephalotes* | Attacep1.0 | Genbank | 10.1371/journal.pgen.1002007 | 748 |
| *Camponotus floridanus* | CamFlo_1.0 | Genbank | 10.1126/science.1192428 | 767 |
| *Cerapachys biroi* | CerBir1.0 | Genbank | 10.1016/j.cub.2014.01.018 | 768 |
| *Ceratosolen solmsi* | CerSol_1.0 | Genbank | 10.1186/gb-2013-14-12-r141 | 897 |
| *Harpegnathos saltator* | HarSal_1.0 | Genbank | 10.1126/science.1192428 | 763 |
| *Lasioglossum albipes* | ASM34657v1 | Genbank | 10.1186/gb-2013-14-12-r142 | 779 |
| *Linepithema humile* | Lhum_1.0 | hymenopteragenome.org | 10.1073/pnas.1008617108 | 762 |
| *Nasonia giraulti* | Ngir_1.0 | Genbank | 10.1126/science.1178028 | 1191 |
| *Nasonia longicornis* | Nlon_1.0 | Genbank | 10.1126/science.1178028 | 1192 |
| *Nasonia vitripennis* | Nvit_2.0 | Genbank | 10.1126/science.1178028 | 1214 |
| *Pogonomyrmex barbatus* | Pbar_UMD_V03 | Genbank | 10.1073/pnas.1007901108 | 666 |
| *Solenopsis invicta* | Si_gnG | Genbank | 10.1073/pnas.1009690108 | 768 |

**Supplemental Table 3**: Quantitative PCR primers used for assessment (relative quantification) of enrichment success and enrichment differences of Cot-1 sources.

| UCE locus | UCE locus genomic position | UCE locus size | Upper primer (5' - 3') | Tm upper primer | Lower primer (5' - 3') | Tm lower primer | Estimated product size |
|---|---|---|---|---|---|---|---|
| uce-82 | chr1:2966279-2966458 | 114 | GCCGACCCCCTGCTGAAGAG | 59.1 | AGACTTACGGCGTCTGCCACG | 59.2 | 77 |
| uce-202 | chr2:4442225-4442404 | 170 | GCCATGCGTGTTCGCTCTTGC | 59.9 | TGCATCGGCCCTTGACAGCG | 60 | 162 |
| uce-591 | chr2:34873617-34873796 | 136 | GGGCATCTACACATTTGAGTCCGCC | 59.9 | ACGAAGTCGAGCCAATTCCATGC | 58 | 102 |
| uce-1101 | chr4:34336396-34336576 | 127 | CGTAGCCATAACGATCGGTCGCC | 59.8 | ACACACCACTGTCGGACAAACTGC | 59.8 | 87 |
| uce-1160 | chr4:5470676-5470856 | 125 | AGGCTTTGGGTGGGCGTTCG | 59.9 | TCACAGCACACACTGGGCCG | 59.6 | 121 |
| uce-1196 | chr4:4001320-4001500 | 137 | GATTAGGGTTGGGGCCTAGGACAGG | 59.8 | GGGGGACAGTACGTGGCTCG | 58.9 | 75 |
| uce-1481 | ChrUn.Scaffold477:51625-51805 | 119 | TCTTCTGCATGGCGTGGTTGG | 57.7 | ACAAGTGCGCTTGCAATTTGTTGGG | 57.7 | 75 |

**Supplemental Table 4**: Crossing point ($C_p$) values for quantitative PCR showing the fold enrichment differences between unenriched controls, enrichments using chicken Cot-1 as a blocking agent, enrichments using hymenoptera Cot-1 as a blocking agent, and Δ Cot-1 or the fold-enrichment difference between chicken and hymenoptera Cot-1.

| | Unenriched controls | | Used chicken Cot-1 | | | | Used hymenoptera Cot-1 | | | | Δ Cot-1 |
|---|---|---|---|---|---|---|---|---|---|---|---|
| | Name | Cp | Name | Cp | Delta Cp | Enrichment | Name | Cp | Delta Cp | Enrichment | |
| Pool 1 | c1-1162 | 21.23 | p1c-1162 | 13.9 | 7.3 | 158.7 | p1h-1162 | 14.4 | 6.8 | 51.0 | 107.6 |
| | c1-132 | 27.27 | p1c-132 | 17.9 | 9.3 | 643.6 | p1h-132 | 18.2 | 9.1 | 184.6 | 459.0 |
| | c1-2055 | 21.82 | p1c-2055 | 13.7 | 8.1 | 280.1 | p1h-2055 | 14.3 | 7.6 | 78.6 | 201.5 |
| | c1-2118 | 21.27 | p1c-2118 | 16.5 | 4.8 | 27.3 | p1h-2118 | 16.9 | 4.4 | 12.6 | 14.6 |
| | c1-2173 | 21.47 | p1c-2173 | 13.9 | 7.5 | 184.8 | p1h-2173 | 14.8 | 6.7 | 46.5 | 138.3 |
| | c1-2704 | 22.10 | p1c-2704 | 14.3 | 7.8 | 227.5 | p1h-2704 | 15.0 | 7.1 | 59.6 | 167.9 |
| | c1-539 | 27.78 | p1c-539 | 18.6 | 9.2 | 580.0 | p1h-539 | 19.6 | 8.2 | 110.5 | 469.5 |
| | c1-neg | | p1c-neg | | | | p1h-neg | | | | |
| Pool 2 | c2-1162 | 22.41 | p2c-1162 | 13.6 | 8.8 | 455.1 | p2h-1162 | 14.3 | 8.1 | 108.0 | 347.1 |
| | c2-1162-r2 | 22.26 | p2c-1162-r2 | 13.4 | 8.8 | 458.3 | p2h-1162-r2 | 14.0 | 8.2 | 114.4 | 343.8 |
| | c2-132 | 27.56 | p2c-132 | 17.5 | 10.1 | 1067.5 | p2h-132 | 18.7 | 8.9 | 168.4 | 899.1 |
| | c2-132-r2 | 27.29 | p2c-132-r2 | 17.3 | 10.0 | 1031.1 | p2h-132-r2 | 17.5 | 9.8 | 284.5 | 746.6 |
| | c2-2055 | 23.23 | p2c-2055 | 13.5 | 9.7 | 855.1 | p2h-2055 | 14.0 | 9.2 | 200.2 | 655.0 |
| | c2-2055-r2 | 22.79 | p2c-2055-r2 | 13.4 | 9.4 | 661.7 | p2h-2055-r2 | 14.0 | 8.8 | 163.6 | 498.1 |
| | c2-2118 | 23.51 | p2c-2118 | 17.1 | 6.4 | 86.8 | p2h-2118 | 17.8 | 5.7 | 27.2 | 59.6 |
| | c2-2118-r2 | 23.22 | p2c-2118-r2 | 17.0 | 6.3 | 76.6 | p2h-2118-r2 | 17.6 | 5.7 | 26.0 | 50.6 |
| | c2-2173 | 23.76 | p2c-2173 | 14.6 | 9.1 | 564.2 | p2h-2173 | 15.3 | 8.5 | 132.1 | 432.0 |
| | c2-2173-r2 | 23.56 | p2c-2173-r2 | 14.4 | 9.1 | 560.3 | p2h-2173-r2 | 15.0 | 8.5 | 136.0 | 424.3 |
| | c2-2704 | 23.52 | p2c-2704 | 14.5 | 9.0 | 515.6 | p2h-2704 | 15.3 | 8.2 | 114.4 | 401.2 |
| | c2-2704-r2 | 23.55 | p2c-2704-r2 | 14.5 | 9.0 | 519.1 | p2h-2704-r2 | 14.9 | 8.6 | 144.9 | 374.2 |
| | c2-539 | 31.04 | p2c-539 | 20.0 | 11.0 | 2091.0 | p2h-539 | 21.0 | 10.1 | 336.3 | 1754.7 |
| | c2-539-r2 | 31.07 | p2c-539-r2 | 19.6 | 11.5 | 2916.5 | p2h-539-r2 | 20.1 | 10.9 | 549.0 | 2367.4 |
| | c2-neg | | p2c-neg | | | | p2h-neg | | | | |
| | c2-neg-r2 | | p2c-neg-r2 | | | | p2h-neg-r2 | | | | |

| | | | | | | | | | | | | | |
|---|---|---|---|---|---|---|---|---|---|---|---|---|---|
| **Pool 3** | c3-1162 | 22.90 | p3c-1162 | 14.0 | 8.9 | 474.4 | p3h-1162 | 14.3 | 8.6 | 145.8 | | 328.7 |
| | c3-132 | 28.58 | p3c-132 | 17.7 | 10.9 | 1910.9 | p3h-132 | 18.0 | 10.6 | 448.7 | | 1462.2 |
| | c3-2055 | 23.29 | p3c-2055 | 13.7 | 9.6 | 792.4 | p3h-2055 | 13.8 | 9.5 | 240.7 | | 551.6 |
| | c3-2118 | 23.18 | p3c-2118 | 16.0 | 7.2 | 149.1 | p3h-2118 | 16.6 | 6.6 | 43.9 | | 105.2 |
| | c3-2173 | 23.70 | p3c-2173 | 14.1 | 9.6 | 797.9 | p3h-2173 | 14.3 | 9.5 | 232.5 | | 565.3 |
| | c3-2704 | 23.14 | p3c-2704 | 13.8 | 9.3 | 643.6 | p3h-2704 | 14.1 | 9.1 | 185.7 | | 457.9 |
| | c3-539 | 30.72 | p3c-539 | 19.7 | 11.0 | 2105.6 | p3h-539 | 19.5 | 11.2 | 641.5 | | 1464.1 |
| | c3-neg | | p3c-neg | | | | p3h-neg | | | | | |
| | | | | | **Avg.** | 744.1 | | | **Avg.** | 178.1 | **Avg.** | 566.0 |
| | | | | | **95 CI** | 259.9 | | | **95 CI** | 57.1 | **95 CI** | 207.0 |

**Supplemental Table 5**: Summary values describing the number of reads collected during sequencing of each enriched library.

| Taxon | Trimmed reads | Total BP | Mean length | 95 % CI | Min lengths | Max length | Median length |
|---|---|---|---|---|---|---|---|
| *Acordulecera pellucida* | 408,901 | 86,125,439 | 210.6 | 0.1 | 40 | 251 | 250 |
| *Andrena (Callandrena) asteris* | 83,975 | 15,540,798 | 185.1 | 0.2 | 40 | 251 | 199 |
| *Andrena (Melandrena) sp* | 410,453 | 84,592,764 | 206.1 | 0.1 | 40 | 251 | 232 |
| *Aphaenogaster albisetosa* | 2,217,687 | 426,328,391 | 192.2 | 0.0 | 40 | 251 | 207 |
| *Aphaenogaster megommata* | 2,047,669 | 386,980,038 | 189.0 | 0.0 | 40 | 251 | 198 |
| *Aphaenogaster tennesseensis* | 1,625,068 | 281,397,272 | 173.2 | 0.0 | 40 | 251 | 173 |
| *Aphaenogaster texana* | 1,059,887 | 182,422,784 | 172.1 | 0.1 | 40 | 251 | 171 |
| *Aporus niger* | 398,607 | 74,624,652 | 187.2 | 0.1 | 40 | 251 | 195 |
| *Bombus pensylvanicus* | 301,910 | 63,481,685 | 210.3 | 0.1 | 40 | 251 | 250 |
| *Chalybion californicus* | 654,184 | 117,929,426 | 180.3 | 0.1 | 40 | 251 | 183 |
| *Chyphotes mellipes* | 1,664,263 | 322,103,690 | 193.5 | 0.0 | 40 | 251 | 208 |
| *Evaniella semaeoda* | 414,086 | 78,104,105 | 188.6 | 0.1 | 40 | 251 | 202 |
| *Messor piceus* | 1,710,354 | 331,117,936 | 193.6 | 0.0 | 40 | 251 | 209 |
| *Metapolybia cingulata* | 719,460 | 142,797,714 | 198.5 | 0.1 | 40 | 251 | 220 |
| *Mischocyttarus flavitarsis* | 307,969 | 61,394,499 | 199.4 | 0.1 | 40 | 251 | 223 |
| *Nasonia vitripennis* | 528,367 | 99,597,773 | 188.5 | 0.1 | 40 | 251 | 199 |
| *Nematus tibialis* | 703,569 | 135,792,550 | 193.0 | 0.1 | 40 | 251 | 213 |
| *Orthogonalys pulchella* | 1,822,967 | 354,456,435 | 194.4 | 0.0 | 40 | 251 | 214 |
| *Pogonomyrmex occidentalis* | 2,129,915 | 406,383,752 | 190.8 | 0.0 | 40 | 251 | 203 |
| *Sapyga pumila* | 1,732,085 | 311,775,579 | 180.0 | 0.0 | 40 | 251 | 180 |
| *Scolia verticalis* | 907,356 | 178,554,253 | 196.8 | 0.1 | 40 | 251 | 221 |
| *Sericomyrmex harekulli* | 327,399 | 64,865,315 | 198.1 | 0.1 | 40 | 251 | 214 |
| *Stenamma diecki* | 1,579,469 | 314,662,462 | 199.2 | 0.0 | 40 | 251 | 218 |
| *Stenamma expolitum* | 1,847,383 | 362,261,429 | 196.1 | 0.0 | 40 | 251 | 211 |
| *Stenamma felixi* | 2,001,433 | 356,384,998 | 178.1 | 0.0 | 40 | 251 | 179 |
| *Stenamma impar* | 1,541,096 | 293,383,544 | 190.4 | 0.0 | 40 | 251 | 199 |
| *Stenamma megamanni* | 2,179,975 | 395,476,057 | 181.4 | 0.0 | 40 | 251 | 181 |
| *Stenamma megamanni2* | 801,435 | 169,627,489 | 211.7 | 0.1 | 40 | 251 | 250 |
| *Stenamma muralla* | 1,237,264 | 238,365,794 | 192.7 | 0.1 | 40 | 251 | 203 |
| *Taxonus pallidicornis* | 577,999 | 119,464,673 | 206.7 | 0.1 | 40 | 251 | 250 |

**Supplemental Table 6**: Summary values describing the number of contigs assembled by ABySS from adapter- and quality-trimmed reads (“All” contigs), their average coverage, the mean length of All contigs, the count of unique reads aligned to All contigs, the number of UCE contigs identified from the pool of All contigs, the mean length of UCE contigs, the average UCE contig sequencing coverage, and the percentage of unique reads that aligned to UCE contigs (this is a percentage of the percentage of unique reads aligning to All contigs).

| Taxon | All contigs | All contigs coverage | All contigs coverage 95 CI | All contigs mean length | All contigs mean length 95 CI | All contigs unique reads aligned | UCE contigs | UCE contigs mean length | UCE contigs coverage | UCE contigs unique reads aligned |
|---|---|---|---|---|---|---|---|---|---|---|
| *Acordulecera pellucida* | 62,419 | 3.4 | 0.1 | 197.4 | 0.8 | 85.6% | 319 | 705.0 | 30.3 | 12.1% |
| *Andrena (Callandrena) asteris* | 9,027 | 4.6 | 0.3 | 208.6 | 2.4 | 79.3% | 714 | 437.1 | 11.5 | 36.9% |
| *Andrena (Melandrena) sp* | 69,660 | 3.3 | 0.2 | 194.1 | 0.7 | 88.2% | 704 | 636.9 | 19.9 | 17.4% |
| *Aphaenogaster albisetosa* | 275,539 | 4.1 | 0.0 | 218.5 | 0.4 | 87.4% | 302 | 654.7 | 43.9 | 2.8% |
| *Aphaenogaster megommata* | 230,940 | 4.1 | 0.1 | 201.2 | 0.4 | 81.0% | 323 | 725.1 | 46.8 | 4.3% |
| *Aphaenogaster tennesseensis* | 184,108 | 4.2 | 0.1 | 186.7 | 0.4 | 82.4% | 412 | 678.6 | 44.9 | 7.2% |
| *Aphaenogaster texana* | 126,796 | 4.0 | 0.1 | 179.4 | 0.4 | 82.6% | 348 | 522.8 | 30.3 | 5.1% |
| *Aporus niger* | 37,593 | 5.0 | 1.0 | 193.1 | 1.0 | 84.1% | 725 | 559.6 | 17.1 | 18.5% |
| *Bombus pensylvanicus* | 55,323 | 3.0 | 0.1 | 197.4 | 0.8 | 88.9% | 703 | 632.0 | 23.9 | 23.7% |
| *Chalybion californicus* | 91,078 | 3.5 | 0.2 | 196.6 | 0.6 | 80.1% | 660 | 614.1 | 34.1 | 18.1% |
| *Chyphotes mellipes* | 191,326 | 4.4 | 0.3 | 222.8 | 0.6 | 89.5% | 472 | 808.1 | 59.7 | 9.2% |
| *Evaniella semaeoda* | 43,255 | 4.4 | 0.2 | 193.8 | 1.1 | 85.9% | 515 | 702.8 | 30.9 | 22.1% |
| *Messor piceus* | 190,453 | 4.5 | 0.2 | 194.0 | 0.4 | 82.8% | 423 | 693.8 | 43.7 | 6.5% |
| *Metapolybia cingulata* | 131,497 | 3.1 | 0.0 | 197.7 | 0.5 | 88.3% | 562 | 618.4 | 38.2 | 13.2% |
| *Mischocyttarus flavitarsis* | 37,614 | 4.4 | 0.3 | 194.0 | 1.0 | 88.1% | 616 | 519.3 | 33.0 | 24.5% |
| *Nasonia vitripennis* | 69,994 | 3.8 | 0.1 | 180.0 | 0.6 | 87.0% | 756 | 463.1 | 45.5 | 22.8% |
| *Nematus tibialis* | 101,439 | 3.4 | 0.1 | 197.3 | 0.7 | 85.4% | 324 | 683.6 | 47.1 | 10.6% |
| *Orthogonalys pulchella* | 174,934 | 3.9 | 0.1 | 251.1 | 0.8 | 83.8% | 266 | 807.8 | 74.1 | 4.6% |
| *Pogonomyrmex occidentalis* | 266,414 | 3.9 | 0.0 | 220.1 | 0.4 | 86.2% | 293 | 658.2 | 56.6 | 3.1% |
| *Sapyga pumila* | 215,619 | 3.7 | 0.1 | 210.1 | 0.5 | 84.4% | 349 | 596.3 | 71.1 | 5.7% |
| *Scolia verticalis* | 105,754 | 3.9 | 0.1 | 208.1 | 0.7 | 84.4% | 516 | 794.1 | 49.2 | 16.2% |
| *Sericomyrmex harekulli* | 61,272 | 3.1 | 0.0 | 185.4 | 0.8 | 89.4% | 663 | 606.1 | 24.6 | 22.1% |
| *Stenamma diecki* | 194,733 | 4.0 | 0.1 | 207.0 | 0.5 | 85.5% | 523 | 820.2 | 46.6 | 10.0% |
| *Stenamma expolitum* | 236,274 | 3.9 | 0.0 | 214.7 | 0.5 | 86.9% | 376 | 785.6 | 50.3 | 5.4% |
| *Stenamma felixi* | 277,927 | 3.5 | 0.0 | 203.8 | 0.4 | 85.1% | 263 | 603.4 | 35.6 | 2.2% |
| *Stenamma impar* | 180,786 | 4.3 | 0.1 | 197.9 | 0.5 | 86.6% | 524 | 727.8 | 45.4 | 9.6% |
| *Stenamma megamanni* | 140,321 | 3.0 | 0.0 | 209.4 | 0.6 | 87.9% | 544 | 803.3 | 35.1 | 12.9% |
| *Stenamma megamanni2* | 285,366 | 3.5 | 0.0 | 213.0 | 0.4 | 86.3% | 314 | 765.0 | 45.6 | 3.8% |
| *Stenamma muralla* | 188,981 | 3.2 | 0.0 | 205.8 | 0.4 | 85.1% | 392 | 713.3 | 46.6 | 7.5% |
| *Taxonus pallidicornis* | 79,453 | 3.5 | 0.1 | 202.9 | 0.8 | 82.5% | 416 | 737.2 | 40.1 | 15.7% |

**Supplemental Table 7**: Summary values describing attributes of the UCE contigs assembled by Trinity.

| Taxon | UCE contigs | UCE contigs total BP | UCE contigs mean length | UCE contigs mean length 95 CI | UCE contigs min length | UCE contigs max length | UCE contigs median length | UCE contigs > 1kb | UCE contigs coverage | UCE contigs unique reads aligned |
|---|---|---|---|---|---|---|---|---|---|---|
| *Acordulecera pellucida* | 341 | 349,519 | 1,025.0 | 19.7 | 206 | 2,504 | 1,054.0 | 198 | 26.3 | 18.4% |
| *Andrena (Callandrena) asteris* | 740 | 425,271 | 574.7 | 7.0 | 202 | 1,447 | 558.0 | 16 | 9.8 | 44.4% |
| *Andrena (Melandrena) sp* | 774 | 663,345 | 857.0 | 9.9 | 208 | 2,253 | 861.5 | 234 | 18.2 | 25.9% |
| *Aphaenogaster albisetosa* | 764 | 862,282 | 1,128.6 | 20.8 | 223 | 11,435 | 1,124.5 | 461 | 88.3 | 26.0% |
| *Aphaenogaster megommata* | 751 | 889,250 | 1,184.1 | 16.2 | 230 | 2,776 | 1,170.0 | 484 | 79.2 | 28.8% |
| *Aphaenogaster tennesseensis* | 751 | 793,300 | 1,056.3 | 14.8 | 210 | 2,493 | 1,051.0 | 412 | 62.4 | 30.5% |
| *Aphaenogaster texana* | 750 | 693,422 | 924.6 | 12.8 | 207 | 2,645 | 906.0 | 281 | 51.5 | 33.2% |
| *Aporus niger* | 740 | 528,399 | 714.1 | 9.1 | 205 | 1,981 | 717.0 | 71 | 14.5 | 17.7% |
| *Bombus pensylvanicus* | 780 | 670,544 | 859.7 | 9.2 | 206 | 2,036 | 879.0 | 228 | 21.4 | 35.0% |
| *Chalybion californicus* | 778 | 631,930 | 812.2 | 10.4 | 205 | 1,968 | 809.0 | 201 | 33.2 | 29.6% |
| *Chyphotes mellipes* | 774 | 916,382 | 1,184.0 | 13.0 | 294 | 3,189 | 1,185.0 | 558 | 66.2 | 26.6% |
| *Evaniella semaeoda* | 638 | 619,918 | 971.7 | 11.6 | 220 | 2,229 | 978.5 | 301 | 31.1 | 39.1% |
| *Messor piceus* | 730 | 811,543 | 1,111.7 | 15.7 | 210 | 3,730 | 1,119.5 | 441 | 58.9 | 26.2% |
| *Metapolybia cingulata* | 685 | 563,953 | 823.3 | 12.8 | 207 | 2,103 | 801.0 | 211 | 40.1 | 24.7% |
| *Mischocyttarus flavitarsis* | 634 | 450,896 | 711.2 | 11.6 | 203 | 2,687 | 676.5 | 110 | 30.0 | 32.4% |
| *Nasonia vitripennis* | 1,166 | 899,101 | 771.1 | 7.7 | 202 | 1,672 | 763.0 | 237 | 46.9 | 57.1% |
| *Nematus tibialis* | 453 | 475,444 | 1,049.5 | 17.9 | 209 | 3,894 | 1,070.0 | 265 | 47.9 | 26.4% |
| *Orthogonalys pulchella* | 706 | 962,959 | 1,364.0 | 16.5 | 205 | 2,998 | 1,352.0 | 569 | 109.0 | 35.0% |
| *Pogonomyrmex occidentalis* | 741 | 846,554 | 1,142.4 | 16.2 | 231 | 3,190 | 1,124.0 | 457 | 97.5 | 26.8% |
| *Sapyga pumila* | 720 | 753,734 | 1,046.9 | 13.4 | 224 | 2,743 | 1,078.0 | 428 | 86.4 | 28.6% |
| *Scolia verticalis* | 760 | 813,497 | 1,070.4 | 12.2 | 286 | 2,877 | 1,078.0 | 461 | 56.6 | 36.0% |
| *Sericomyrmex harekulli* | 744 | 606,204 | 814.8 | 9.7 | 205 | 2,099 | 830.5 | 177 | 22.3 | 33.5% |
| *Stenamma diecki* | 751 | 857,659 | 1,142.0 | 15.0 | 209 | 3,188 | 1,167.0 | 488 | 53.5 | 23.7% |
| *Stenamma expolitum* | 749 | 907,836 | 1,212.1 | 15.7 | 205 | 2,690 | 1,216.0 | 520 | 69.3 | 25.7% |
| *Stenamma felixi* | 762 | 816,726 | 1,071.8 | 14.2 | 209 | 3,469 | 1,056.0 | 433 | 75.3 | 25.1% |
| *Stenamma impar* | 741 | 782,478 | 1,056.0 | 14.1 | 229 | 2,846 | 1,046.0 | 428 | 49.8 | 22.4% |
| *Stenamma megamanni* | 754 | 858,069 | 1,138.0 | 14.5 | 217 | 3,227 | 1,166.5 | 502 | 37.8 | 28.6% |
| *Stenamma megamanni2* | 756 | 932,105 | 1,232.9 | 20.0 | 204 | 9,956 | 1,218.0 | 525 | 87.5 | 28.7% |
| *Stenamma muralla* | 734 | 830,910 | 1,132.0 | 14.8 | 221 | 3,299 | 1,113.0 | 480 | 61.6 | 30.6% |
| *Taxonus pallidicornis* | 459 | 523,674 | 1,140.9 | 21.6 | 205 | 3,001 | 1,173.0 | 282 | 37.7 | 27.5% |

**Supplemental Table 8**: Summary values describing attributes of the UCE contigs assembled by ABySS.

| Taxon | UCE contigs | UCE contigs total BP | UCE contigs mean length | UCE contigs mean length 95 CI | UCE contigs min length | UCE contigs max length | UCE contigs median length | UCE contigs > 1kb | UCE contigs coverage | UCE contigs unique reads aligned |
|---|---|---|---|---|---|---|---|---|---|---|
| *Acordulecera pellucida* | 319 | 224,885 | 705.0 | 17.0 | 102 | 1,774 | 737.0 | 52 | 30.3 | 12.1% |
| *Andrena (Callandrena) asteris* | 714 | 312,069 | 437.1 | 5.5 | 102 | 901 | 431.0 | 0 | 11.5 | 36.9% |
| *Andrena (Melandrena) sp* | 705 | 449,059 | 637.0 | 9.3 | 103 | 1,577 | 635.0 | 47 | 19.9 | 17.4% |
| *Aphaenogaster albisetosa* | 302 | 197,727 | 654.7 | 19.3 | 105 | 2,072 | 621.0 | 48 | 43.9 | 2.8% |
| *Aphaenogaster megommata* | 323 | 234,220 | 725.1 | 19.5 | 103 | 1,818 | 687.0 | 63 | 46.8 | 4.3% |
| *Aphaenogaster tennesseensis* | 413 | 280,246 | 678.6 | 15.5 | 101 | 1,800 | 647.0 | 71 | 44.9 | 7.2% |
| *Aphaenogaster texana* | 348 | 181,931 | 522.8 | 13.5 | 104 | 1,714 | 496.0 | 12 | 30.3 | 5.1% |
| *Aporus niger* | 726 | 406,631 | 560.1 | 7.6 | 101 | 1,230 | 547.5 | 12 | 17.1 | 18.5% |
| *Bombus pensylvanicus* | 703 | 444,314 | 632.0 | 8.3 | 101 | 1,521 | 635.0 | 26 | 23.9 | 23.7% |
| *Chalybion californicus* | 660 | 405,336 | 614.1 | 9.5 | 101 | 1,336 | 616.0 | 34 | 34.1 | 18.1% |
| *Chyphotes mellipes* | 472 | 381,416 | 808.1 | 17.1 | 102 | 2,063 | 829.5 | 140 | 59.7 | 9.2% |
| *Evaniella semaeoda* | 515 | 361,966 | 702.8 | 12.1 | 104 | 2,191 | 721.0 | 61 | 30.9 | 22.1% |
| *Messor piceus* | 424 | 294,305 | 694.1 | 16.1 | 101 | 2,044 | 669.5 | 62 | 43.7 | 6.5% |
| *Metapolybia cingulata* | 563 | 348,507 | 619.0 | 12.4 | 101 | 1,677 | 604.0 | 62 | 38.2 | 13.2% |
| *Mischocyttarus flavitarsis* | 617 | 320,960 | 520.2 | 9.3 | 101 | 1,878 | 503.0 | 13 | 33.0 | 24.5% |
| *Nasonia vitripennis* | 756 | 350,095 | 463.1 | 8.7 | 101 | 1,368 | 450.5 | 17 | 45.5 | 22.8% |
| *Nematus tibialis* | 325 | 222,333 | 684.1 | 18.4 | 101 | 1,972 | 713.0 | 53 | 47.1 | 10.6% |
| *Orthogonalys pulchella* | 266 | 214,864 | 807.8 | 28.4 | 102 | 2,303 | 826.5 | 96 | 74.1 | 4.6% |
| *Pogonomyrmex occidentalis* | 293 | 192,855 | 658.2 | 19.9 | 106 | 2,134 | 632.0 | 36 | 56.6 | 3.1% |
| *Sapyga pumila* | 349 | 208,097 | 596.3 | 16.5 | 102 | 1,533 | 555.0 | 35 | 71.1 | 5.7% |
| *Scolia verticalis* | 516 | 409,772 | 794.1 | 13.8 | 101 | 1,857 | 803.0 | 137 | 49.2 | 16.2% |
| *Sericomyrmex harekulli* | 663 | 401,814 | 606.1 | 8.8 | 101 | 1,613 | 618.0 | 22 | 24.6 | 22.1% |
| *Stenamma diecki* | 524 | 429,972 | 820.6 | 15.2 | 101 | 1,850 | 818.0 | 164 | 46.6 | 10.0% |
| *Stenamma expolitum* | 376 | 295,389 | 785.6 | 18.5 | 101 | 2,136 | 764.0 | 95 | 50.3 | 5.4% |
| *Stenamma felixi* | 263 | 158,700 | 603.4 | 17.4 | 103 | 1,843 | 558.0 | 21 | 35.6 | 2.2% |
| *Stenamma impar* | 524 | 381,375 | 727.8 | 13.7 | 102 | 2,480 | 690.5 | 107 | 45.4 | 9.6% |
| *Stenamma megamanni2* | 314 | 240,207 | 765.0 | 20.2 | 102 | 1,980 | 719.5 | 67 | 45.6 | 3.8% |
| *Stenamma megamanni* | 545 | 437,935 | 803.6 | 15.5 | 101 | 2,653 | 793.0 | 164 | 35.1 | 12.9% |
| *Stenamma muralla* | 392 | 279,628 | 713.3 | 16.1 | 102 | 2,114 | 664.0 | 70 | 46.6 | 7.5% |
| *Taxonus pallidicornis* | 416 | 306,665 | 737.2 | 18.8 | 102 | 2,906 | 724.5 | 100 | 40.1 | 15.7% |

**Supplemental Table 9**: Model structure, AIC, number of parameters, AICc, and Akaike weight ($w_i$) for general linear models of parameters affecting the mean number of UCE contigs captured.

| | model | AIC | Params | AICc | $\Delta_i$ | $w_i$ |
|---|---|---|---|---|---|---|
| 1 | contigs ~ distance + reads + assembly + mean | 1241.0 | 6 | 1244.7 | 0.0 | 1.0 |
| 2 | contigs ~ distance + reads + mean | 1250.6 | 5 | 1253.1 | 8.4 | 0.0 |
| 3 | contigs ~ distance + assembly + mean | 1579.9 | 5 | 1582.4 | 337.7 | 0.0 |
| 4 | contigs ~ distance + mean | 1683.0 | 4 | 1684.6 | 439.9 | 0.0 |
| 5 | contigs ~ distance + reads + assembly | 1744.7 | 5 | 1747.2 | 502.5 | 0.0 |
| 6 | contigs ~ distance + assembly | 1755.0 | 4 | 1756.6 | 511.9 | 0.0 |
| 7 | contigs ~ reads + assembly + mean | 2593.5 | 5 | 2596.0 | 1351.3 | 0.0 |
| 8 | contigs ~ reads + assembly | 2609.5 | 4 | 2611.1 | 1366.4 | 0.0 |
| 9 | contigs ~ assembly + mean | 2629.4 | 4 | 2631.0 | 1386.3 | 0.0 |
| 10 | contigs ~ assembly | 2757.2 | 3 | 2758.1 | 1513.5 | 0.0 |
| 11 | contigs ~ distance + reads | 3147.4 | 4 | 3149.0 | 1904.3 | 0.0 |
| 12 | contigs ~ distance | 3157.7 | 3 | 3158.6 | 1914.0 | 0.0 |
| 13 | contigs ~ reads + mean | 3235.8 | 4 | 3237.4 | 1992.7 | 0.0 |
| 14 | contigs ~ mean | 3801.1 | 3 | 3802.0 | 2557.4 | 0.0 |
| 15 | contigs ~ reads | 4100.1 | 3 | 4101.0 | 2856.4 | 0.0 |

**Supplemental Table 10**: Model structure, AIC, number of parameters, AICc, and Akaike weight ($w_i$) for general linear models of parameters affecting the number of UCE contigs captured among Trinity (only) assemblies.

| | **model** | **AIC** | **Params** | **AICc** | $\Delta_i$ | $w_i$ |
|---|---|---|---|---|---|---|
| 1 | contigs ~ distance + mean | 424.2 | 3 | 425.2 | 0.0 | 0.8 |
| 2 | contigs ~ distance + reads + mean | 426.2 | 4 | 427.8 | 2.6 | 0.2 |
| 3 | contigs ~ distance + reads | 457.3 | 3 | 458.2 | 33.0 | 0.0 |
| 4 | contigs ~ distance | 571.9 | 2 | 572.3 | 147.2 | 0.0 |
| 5 | contigs ~ reads + mean | 829.3 | 3 | 830.3 | 405.1 | 0.0 |
| 6 | contigs ~ reads | 997.8 | 2 | 998.2 | 573.1 | 0.0 |
| 7 | contigs ~ mean | 998.8 | 2 | 999.3 | 574.1 | 0.0 |

**Supplemental Table 11**: Model structure, AIC, number of parameters, AICc, and Akaike weight ($w_i$) for general linear models of parameters affecting the length of UCE contigs captured.

| | **model** | **AIC** | **Params** | **AICc** | **Δi** | $w_i$ |
|---|---|---|---|---|---|---|
| 1 | mean ~ distance + reads + assembly | 687.3 | 5 | 689.8 | 0.0 | 1.0 |
| 2 | mean ~ distance + assembly | 721.6 | 4 | 723.2 | 33.3 | 0.0 |
| 3 | mean ~ reads + assembly | 744.9 | 4 | 746.5 | 56.6 | 0.0 |
| 4 | mean ~ assembly | 775.2 | 3 | 776.1 | 86.3 | 0.0 |
| 5 | mean ~ distance + reads | 781.5 | 4 | 783.1 | 93.3 | 0.0 |
| 6 | mean ~ distance | 788.4 | 3 | 789.3 | 99.5 | 0.0 |
| 7 | mean ~ reads | 814.7 | 3 | 815.6 | 125.8 | 0.0 |

**Supplemental Table 12**: Model structure, AIC, number of parameters, AICc, and Akaike weight ($w_i$) for general linear models of parameters affecting the length of UCE contigs captured among Trinity (only) assemblies.

| | **model** | **AIC** | **Params** | **AICc** | **$\Delta_i$** | **$w_i$** |
|---|---|---|---|---|---|---|
| 1 | contig length ~ distance + reads | 333.1 | 3 | 334.0 | 0.0 | 1.00 |
| 2 | contig length ~ distance | 373.6 | 2 | 374.0 | 40.0 | 0.00 |
| 3 | contig length ~ reads | 375.8 | 2 | 376.3 | 42.3 | 0.00 |

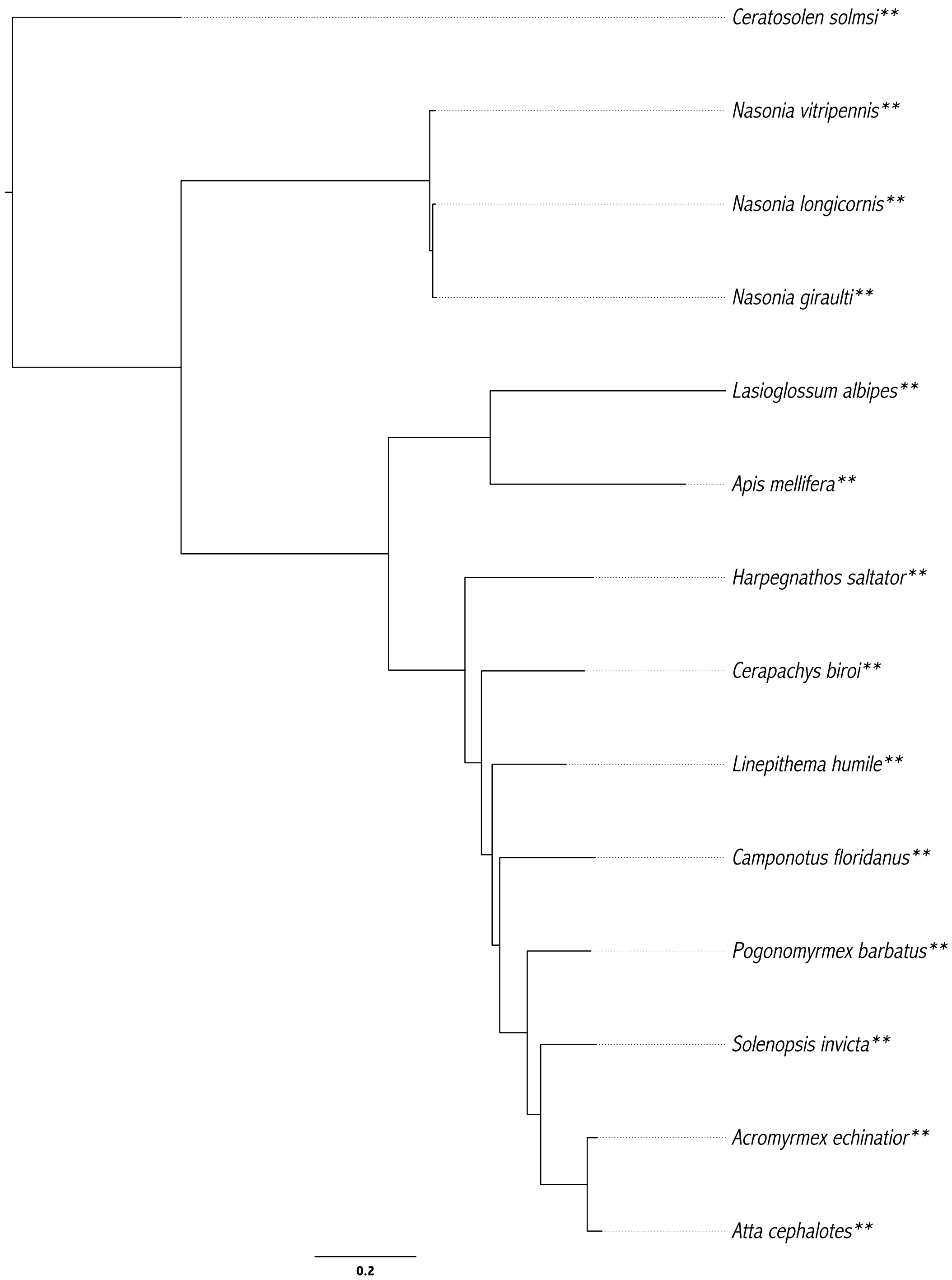

**Supplemental Figure 1**: Maximum likelihood phylogeny inferred from a 75% complete supermatrix containing data from ultraconserved elements identified in 14 genome-enabled taxa. We show bootstrap support values only where support is < 100%. Although genome assemblies exist for additional hymenopteran taxa, we were not granted permission to include these data in our analyses.

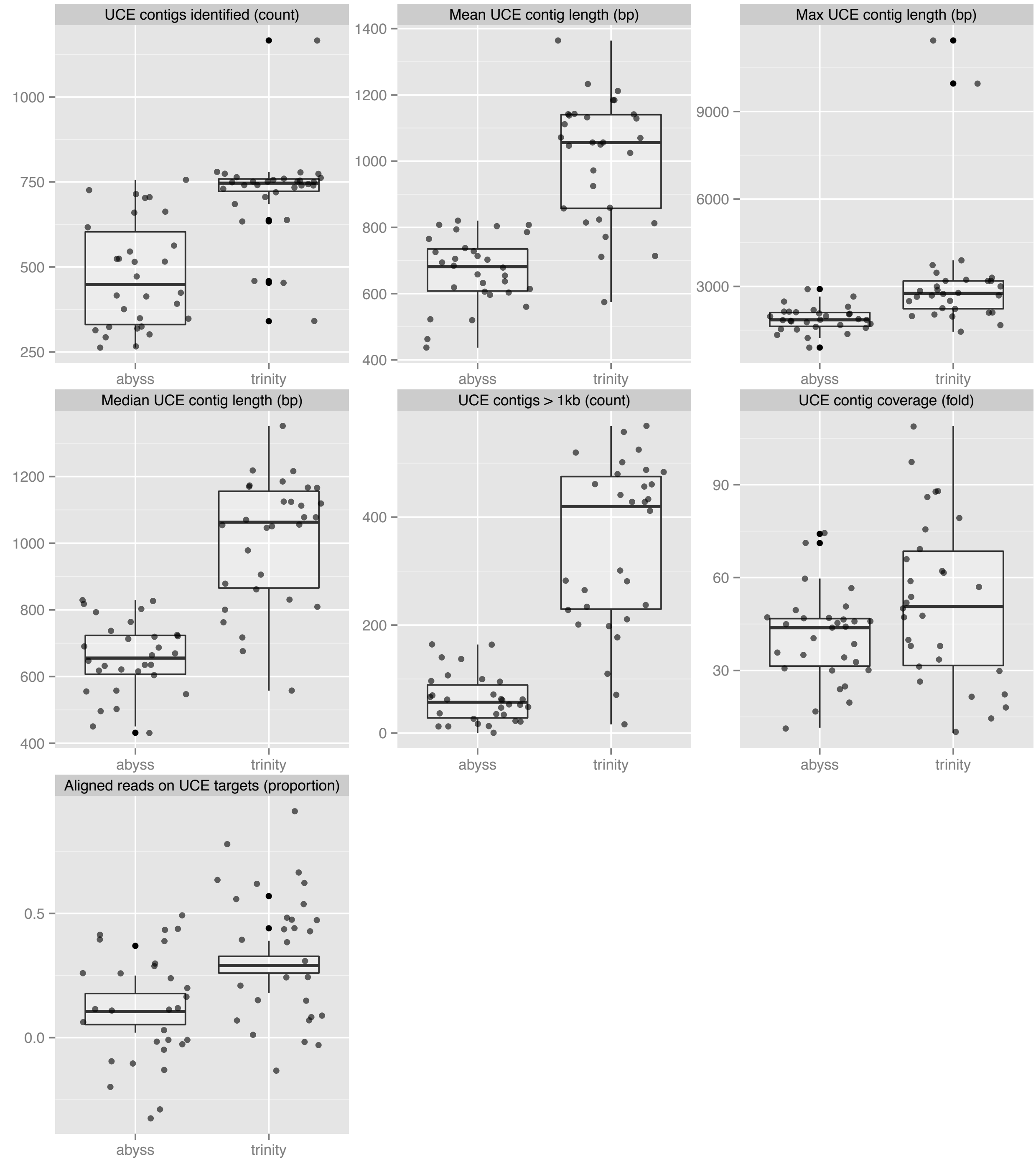

**Supplemental Figure 2**: Box plots showing differences in standard metrics among UCE contigs assembled by Trinity or ABySS. Jittered dots indicate a given value for each taxon. Values correspond to those in Supplemental Table 7 and Supplemental Table 8.

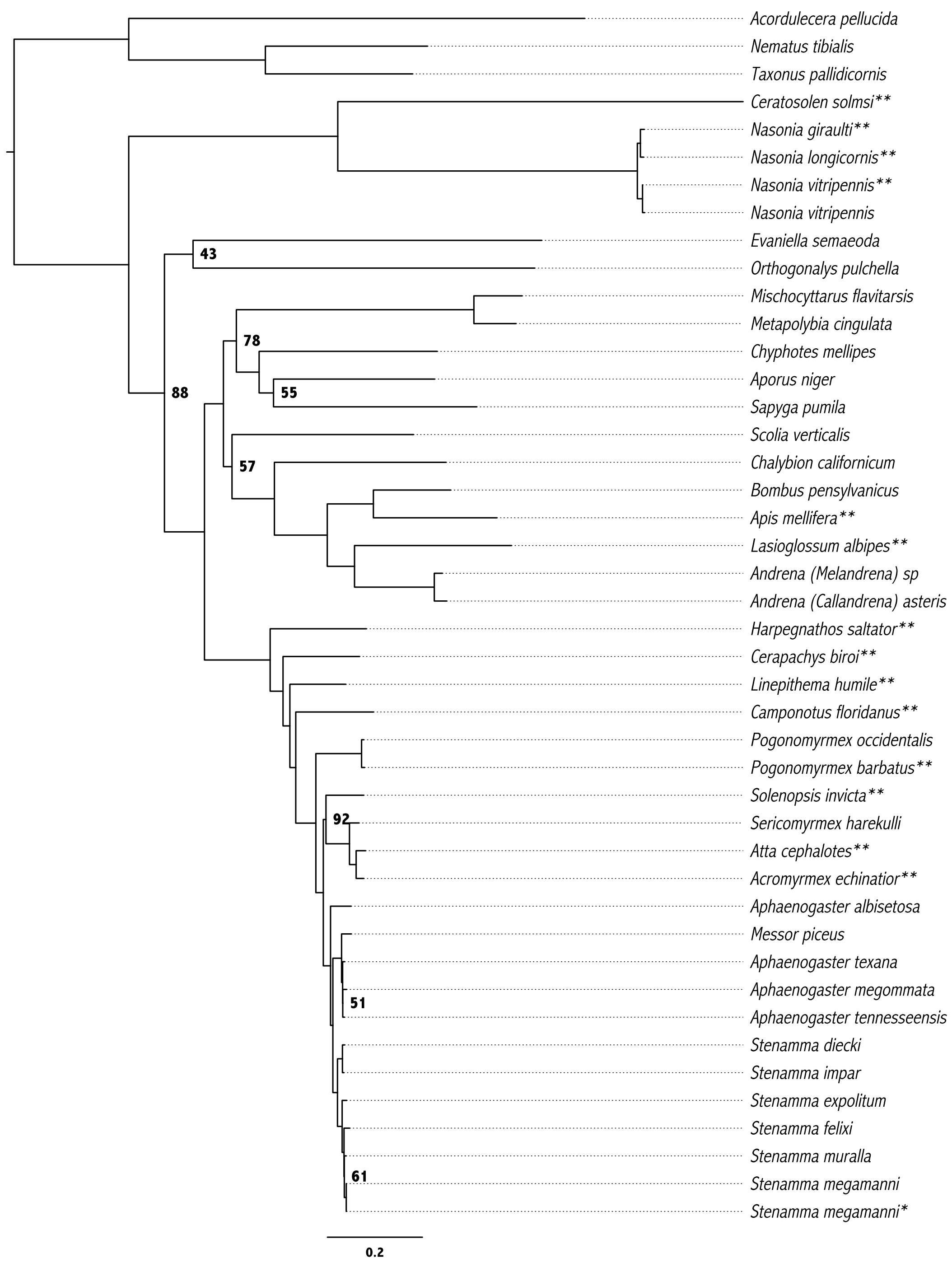

**Supplemental Figure 3**: Maximum likelihood phylogeny inferred from a 75% complete supermatrix containing data from 14 genome-enabled taxa (identified by double-asterisks) and 30 taxa from which we enriched and assembled (ABySS) ultraconserved element loci. We show bootstrap support values only where support is < 100%, and the single asterisk beside *Stenamma megamanni* denotes that this sample represents a different population of the same species.

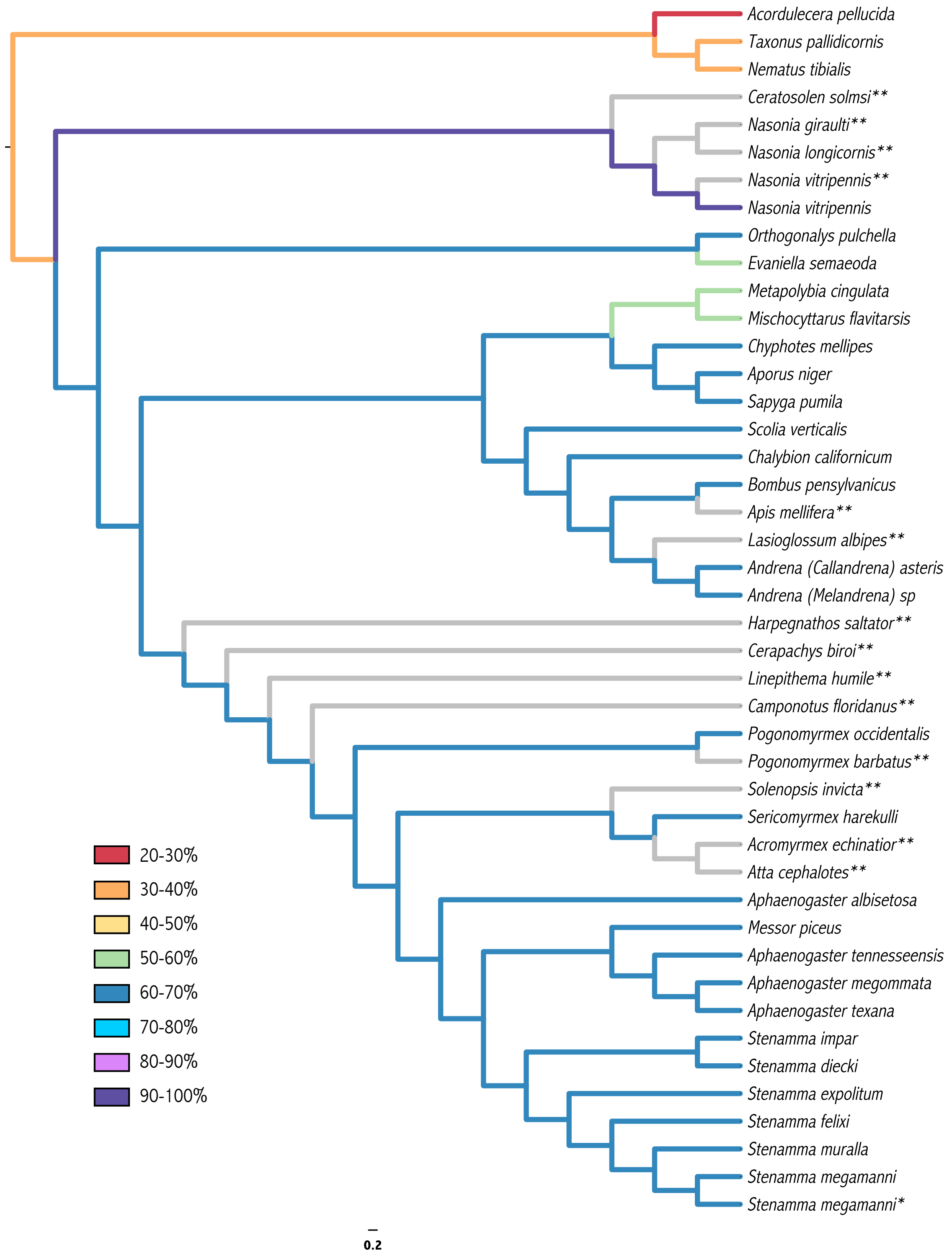

**Supplemental Figure 4**: The topology from Figure 1, with branches colored to indicate the approximate number of ultraconserved element loci we captured, by taxon, relative to the total number of loci captured from *Nasonia vitripennis* (n=1,166).

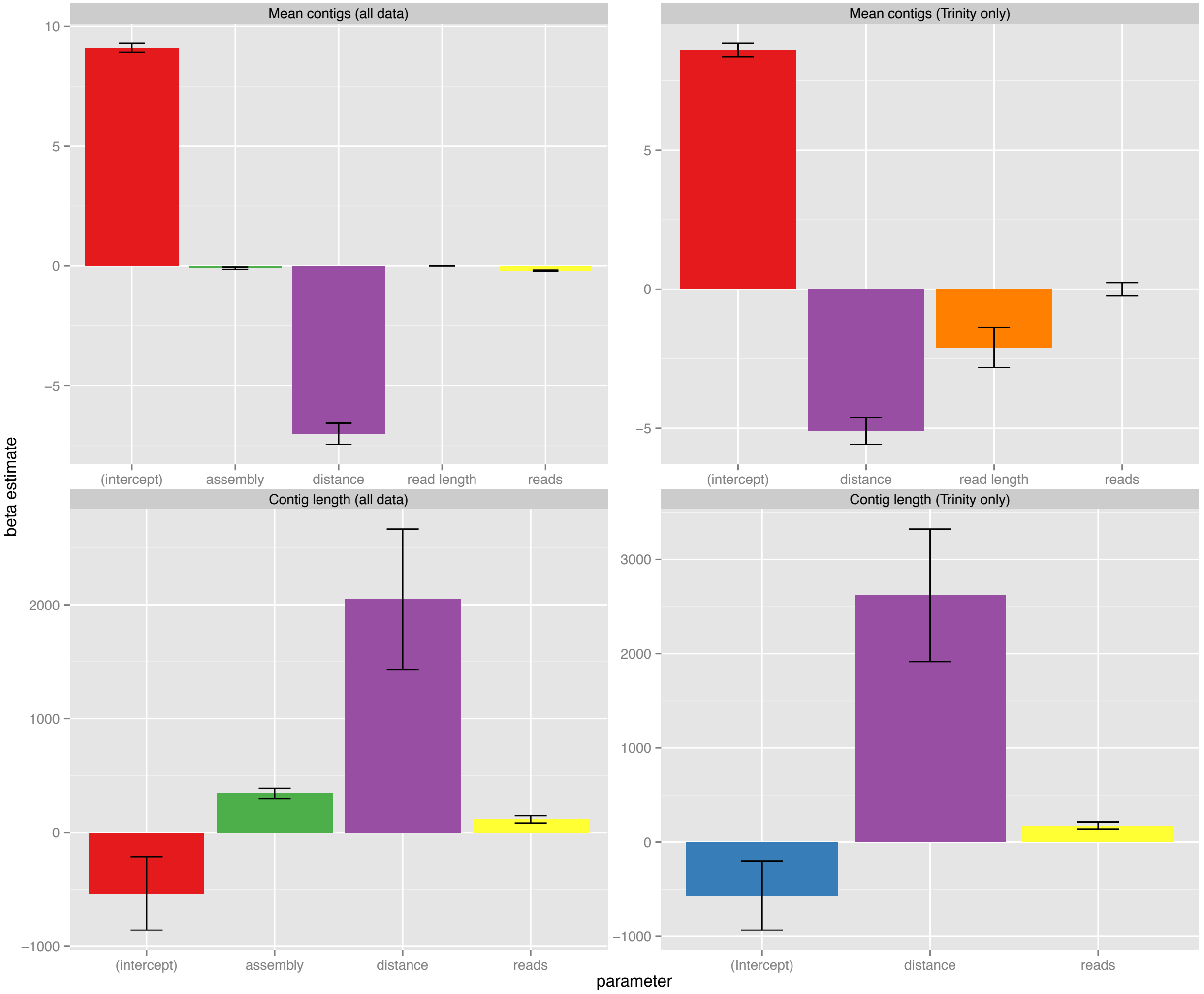

**Supplemental Figure 5**: Bar plots comparing parameter (β) estimates (± 95% CI) from general linear models of factors affecting the number of UCE contigs enriched and the length of enriched UCE contigs. Note that the y-axis differs across sub-panels.

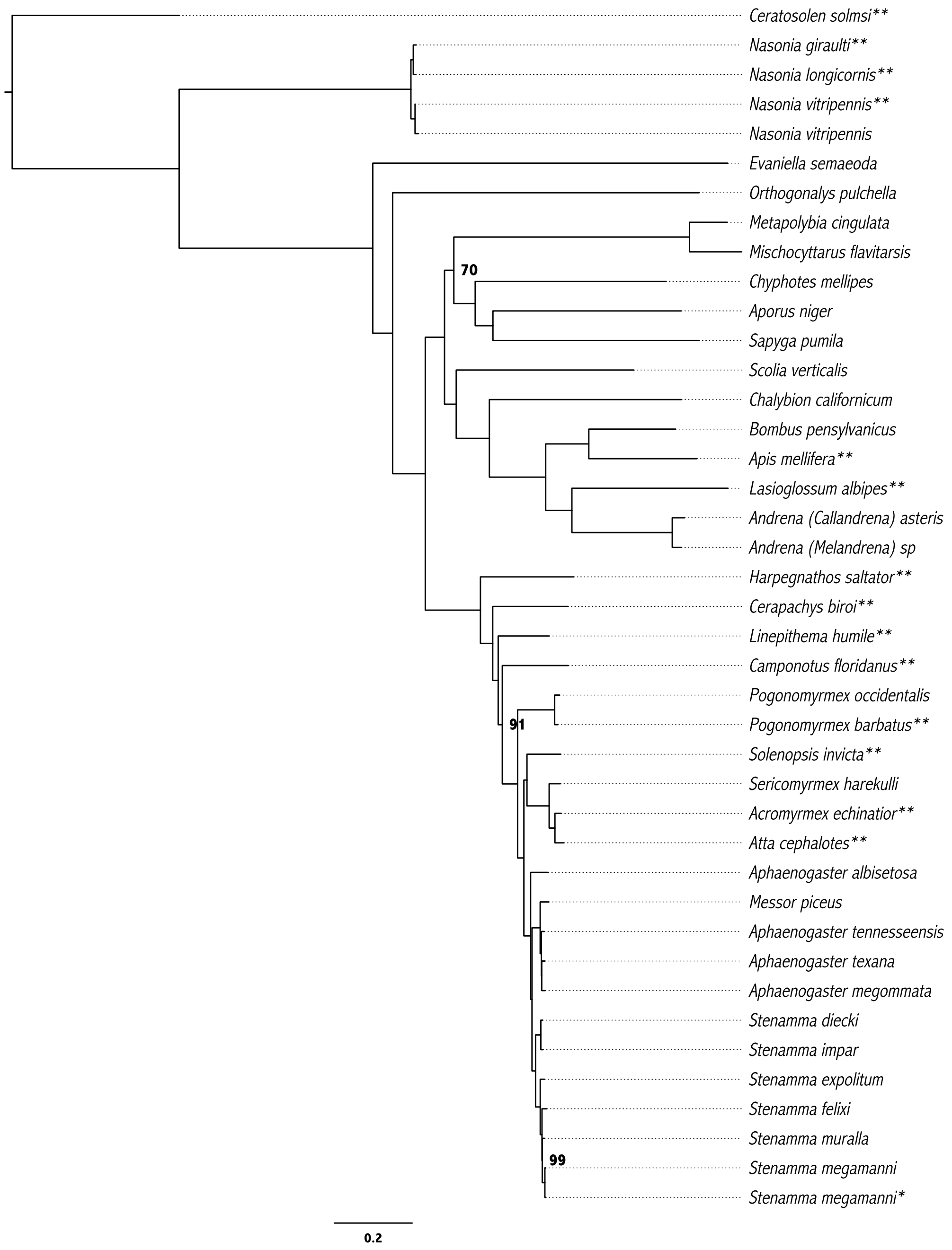

**Supplemental Figure 6**: Maximum likelihood phylogeny inferred from a 75% complete supermatrix containing data from 14 genome-enabled taxa (identified by double-asterisks) and 27 taxa from which we enriched and assembled (Trinity) ultraconserved element loci. To infer this tree, we removed three sawfly taxa from consideration, re-identified UCE loci, re-extracted relevant UCE contigs, and re-aligned the extracted data, resulting in a slightly larger data matrix from that in Figure 1. We show bootstrap support values only where support is < 100%, and the single asterisk beside *Stenamma megamanni* denotes that this sample represents a different population of the same species.